\documentclass[aps,preprint,showpacs,showkeys,nofootinbib,superscriptaddress]{revtex4}

\usepackage{gensymb} 
\usepackage{textcomp}

\usepackage{graphicx}  %
\usepackage{amsmath}
\usepackage{bm}  %

\newcommand{\bea}{\begin{eqnarray}}
\newcommand{\eea}{\end{eqnarray}}
\newcommand{\beq}{\begin{equation}}
\newcommand{\eeq}{\end{equation}}
\newcommand{\bqa}{\begin{eqnarray}}
\newcommand{\eqa}{\end{eqnarray}}

\def\mqo2{{\!\!\!}}

\begin{document}

\title{
Production of the $\bm{X(3872)}$ at the Tevatron and the LHC}
\author{Pierre Artoisenet}
\affiliation{Department of Physics,
         The Ohio State University, Columbus, OH\ 43210, USA\\}
\author{Eric Braaten}
\affiliation{Department of Physics,
         The Ohio State University, Columbus, OH\ 43210, USA\\}
\date{\today}

\begin{abstract}
We predict the differential cross sections for production of the 
$X(3872)$ at the Tevatron and the Large Hadron Collider 
from both prompt QCD mechanisms and from decays of $b$ hadrons.
The prompt cross section is calculated using the NRQCD
factorization formula.  Simplifying assumptions are used to 
reduce the nonperturbative parameters to a single
NRQCD matrix element that is determined from an estimate
of the prompt cross section at the Tevatron.
For $X(3872)$ with transverse momenta greater than about 4~GeV, 
the predicted cross section is insensitive to the simplifying 
assumptions.  
We also discuss critically a recent analysis that concluded that 
the prompt production rate at the Tevatron is too large by 
orders of magnitude for the $X(3872)$ to be a weakly-bound 
charm-meson molecule.  We point out that if charm-meson rescattering
is properly taken into account, the upper bound is increased
by orders of magnitude and is compatible with the 
observed production rate at the Tevatron.
\end{abstract}

\smallskip
\maketitle

\section{Introduction}
\label{sec:intro}

The $X(3872)$ is a $c \bar c$ meson that was discovered by the 
Belle Collaboration in 2003 through 
the decay $B^+ \to K^+ + X$~\cite{Choi:2003ue}.  
Its existence was quickly confirmed by the CDF Collaboration through 
inclusive production of $X(3872)$ in high energy $p \bar p$  
collisions \cite{Acosta:2003zx}.
While some of the $X(3872)$'s at the Tevatron are produced by $b$-hadron
decays, most are produced promptly by QCD mechanisms~\cite{Bauer:2004bc}.
A compelling case can be made that the $X(3872)$
is a loosely-bound charm-meson molecule whose particle content is
\beq
X = \frac{1}{\sqrt2}
\left( D^{*0} \bar D^0 + D^0 \bar D^{*0} \right)
\equiv \left( D^{*0} \bar D^0 \right)_+.
\label{X-DD}
\eeq
It has univeral properties that are determined by its 
binding energy only \cite{Voloshin:2003nt,Braaten:2003he}.
They imply that the mean separation of its constituents could be larger 
than for ordinary hadrons by an order of magnitude or more.  
Given the identification of the $X(3872)$ as a charm-meson molecule, 
an obvious question is whether the $X(3872)$ 
could still be sufficiently robust to be observed at a hadron collider 
like the Tevatron.  This issue has been brought into focus 
by a recent paper by Bignamini et al.\ \cite{Bignamini:2009sk}
that derives an upper bound on the prompt cross section for a 
charm-meson molecule in terms of charm-meson-pair cross sections.  
They apply their upper bound to the $X(3872)$ and conclude that the 
observed prompt production rate at the Tevatron exceeds their bound 
by orders of magnitude.

The key to understanding how the $X(3872)$ evades the upper bound of 
Ref.~\cite{Bignamini:2009sk} is rescattering.  In this paper, 
we will show that if rescattering of the charm mesons is properly 
taken into account, the upper bound of Ref.~\cite{Bignamini:2009sk} 
is increased by orders of magnitude, which brings it into consistency 
with measurements at the Tevatron.  Thus, despite its weak binding, 
the $X(3872)$ is sufficiently robust to be studied at a hadron collider.
Studies by the CDF Collaboration at the Tevatron have provided the 
strongest constraints thus far on the quantum numbers of the 
$X(3872)$ \cite{Abulencia:2006ma} and the most precise measurement 
of its binding energy \cite{Aaltonen:2009vj}.
At the Large Hadron Collider (LHC) at CERN,
the data samples of the $X(3872)$ will be much larger than at the 
Tevatron and thus will allow some of its properties to be studied
with much greater precision.

In this paper, we also predict the differential cross sections 
for production of $X(3872)$ at the Tevatron and the LHC.
We calculate the prompt production rates from QCD mechanisms
using the NRQCD factorization formalism \cite{Braaten:2004jg}, 
which expresses the cross section as the sum of 
parton cross sections for creating 
$c \bar c$ pairs with vanishing relative momentum multiplied 
by phenomenological constants.  
We use simplifying assumptions to reduce the phenomenological constants 
to a single NRQCD matrix element, which is determined using an
estimate of the prompt cross section at the Tevatron.
We also calculate the rates for production of the $X(3872)$ 
from $b$-hadron decays using a method that gives the observed 
production rates of the $J/\psi$ from $b$-hadron decays at the Tevatron.

We begin in Section~\ref{sec:X3872} by
summarizing the case for the $X(3872)$ as a loosely-bound 
charm-meson molecule and describing some of its universal properties.
We also derive a factorization formula that determines the 
dependence of the production rate on the binding energy of the $X(3872)$.
In Section~\ref{sec:CDFdata}, we use results from the CDF Collaboration 
to estimate the cross sections for production of the $X(3872)$
at the Tevatron from both prompt QCD mechanisms and $b$-hadron decays.
In Section~\ref{sec:charmpair}, we consider the relation between 
the production of the $X(3872)$ and 
the production of pairs of charm mesons.
We point out that if the effects of charm-meson rescattering 
are taken into account, the upper bound on the cross section 
for the $X(3872)$ that was derived in Ref.~\cite{Bignamini:2009sk} 
is increased by orders of magnitude.
We also present an order-of-magnitude estimate of the $X(3872)$ cross section 
and show that it is compatible with measurements at the Tevatron.
In Section~\ref{sec:NRQCD}, we use the NRQCD factorization formula 
for inclusive production to calculate the 
differential cross sections for prompt $X(3872)$ at the Tevatron.
We also calculate the differential cross sections for 
$X(3872)$ from $b$-hadron decays at the Tevatron.
In Section~\ref{sec:LHC}, we calculate the differential cross sections for 
$X(3872)$ at the LHC from both prompt QCD mechanisms 
and $b$-hadron decays.  Finally, we summarize our results in 
Section~\ref{sec:summary}.

\section{The $\bm{X(3872)}$}
\label{sec:X3872}

In this section, we summarize the case for the $X(3872)$ as a 
loosely-bound charm-meson molecule whose particle content 
is given in Eq.~(\ref{X-DD}).
We also derive a factorization formula for production rates
of the $X(3872)$ that reveals how the rate depends on its 
binding energy.

\subsection{Universal Properties}

The only experimental information that is necessary to make the 
identification of the $X(3872)$ as a loosely-bound 
charm-meson molecule is the determination of its quantum numbers 
and the measurements of its mass.
The quantum numbers of the $X(3872)$ are $1^{++}$, which follows from
\begin{itemize}
\item
the observation of its decay into $J/\psi \, \gamma$, which implies 
that it is even under charge conjugation \cite{Abe:2005ix,Aubert:2006aj},
\item
analyses of the momentum distributions from its decay into 
$J/\psi \, \pi^+ \pi^-$, which imply that its spin and parity are 
$1^+$ or $2^-$ \cite{Abe:2005iya,Abulencia:2006ma},
\item
either the observation of its decays into 
$D^0 \bar D^0 \pi^0$ \cite{Gokhroo:2006bt},
which disfavors $2^-$ because of angular momentum suppression,
or the observation of its decay into 
$\psi(2S) \, \gamma$ \cite{Babar:2008rn},
which disfavors $2^-$ because of multipole suppression.
\end{itemize}
The most recent measurements of the mass of the $X(3872)$ in the 
$J/\psi \, \pi^+ \pi^-$ decay mode 
\cite{Aubert:2008gu,Belle:2008te,Aaltonen:2009vj}
imply that its energy relative to the $D^{*0} \bar D^0$ threshold is
\beq
M_X - ( M_{D^{*0}} + M_{D^0} ) = -0.30 \pm 0.40 ~ {\rm MeV}.
\label{MX}
\eeq
The quantum numbers $1^{++}$ imply that the $X(3872)$ has an S-wave 
coupling to $D^{*0} \bar D^0$.  Its tiny energy relative to the
$D^{*0} \bar D^0$ threshold implies that it is a resonant coupling.  
Thus the $X(3872)$ is an S-wave threshold resonance.

A system of particles with short-range interactions 
that are fine-tuned so that pairs of particles have
an S-wave threshold resonance 
is predicted by nonrelativistic quantum mechanics 
to have universal behavior~\cite{Braaten:2004rn}.
The universal properties of the system are determined
by the pair scattering length $a$,
which is large compared to the length scale $1/\Lambda$ 
set by the range of the interactions.
The universal elastic scattering amplitude is
\beq
f(E) = \frac{1}{-1/a + \sqrt{-2 \mu E - i \epsilon}},
\label{f-E}
\eeq
where $E$ is the energy relative to the scattering threshold
and $\mu$ is the reduced mass of the pair.
This scattering amplitude is applicable
in the region $|E| < \Lambda^2/(2 \mu)$, 
with errors that scale as $(2 \mu |E|)^{1/2}/\Lambda$.
If $a>0$, one of the universal properties is the existence 
of a molecule with binding energy $E_X = 1/(2 \mu a^2)$.
The universal wavefunction for this molecule is
\beq
\psi_X(\bm{r}) = \frac{1}{\sqrt{2 \pi a} \, r} \exp(-r/a).
\label{psi-uni}
\eeq
The universal wavefunction in the momentum representation is
\beq
\tilde \psi_X ({\bm k})=\frac{\sqrt{8 \pi/a}}{k^2+1/a^2} ,
\label{psi-uni:k}
\eeq
which is applicable for $k < \Lambda$.
The wavefunction in the coordinate representation in 
Eq.~(\ref{psi-uni}) implies that the root-mean-square (rms)
separation of the constituents is $r_X = a/\sqrt2$.

To apply the universal properties of S-wave threshold resonances 
to the $X(3872)$, we take the scattering length $a$ to be that for 
charm mesons in the 
channel $( D^{*0} \bar D^0 )_+$ defined in Eq.~(\ref{X-DD}).
Assuming $a>0$, the scattering length $a = (2 \mu E_X)^{-1/2}$ 
can be determined by using the binding energy from Eq.~(\ref{MX}).
Taking $E_X = 0.30 \pm 0.40$~MeV as the input,
we find that the charm mesons in the $X(3872)$
have an astonishingly large rms separation:
$r_X = 5.8^{+\infty}_{-2.0}$~fm.
A reasonable estimate for the momentum scale $\Lambda$ 
associated with the range of the interactions between 
the charm mesons is the pion mass $m_\pi \approx 135$~MeV.  
The binding energy $E_X$ is very small compared to the 
corresponding energy scale $m_\pi^2/(2 \mu) \approx 9.4$~MeV.

\subsection{Factorization of the Production Rate}
\label{sec:factprod}

Production rates of the $X(3872)$ satisfy a factorization 
formula that is based on separating the momentum scale
$\sqrt{2 \mu E_X}$ associated with the binding of the $X$ from
all the larger momentum scales of QCD \cite{Braaten:2005jj},
which include the pion mass $m_\pi$ and the charm quark mass $m_c$. 
Our starting point for the derivation of the factorization formula 
is the differential cross section for a pair of charm
mesons with energy near the threshold:
\beq
d \sigma[D^{*0} \bar D^0(\bm{k}) ] =
\frac{1}{\rm flux} \sum_{\rm all} \int d \phi_{D^* \bar D + {\rm all}} 
\left| {\cal T}[D^{*0} \bar D^0(\bm{k}) \,+\, {\rm all}] \right|^2  
\frac{d^3k}{(2 \pi)^3 2\mu} ,
\label{sig-above-thres}
\eeq
where $\bm{k}$ is the relative momentum of $D^{*0} \bar D^0$
in the center-of-momentum frame of the pair and ``all'' 
represents all the other possible particles in the final state.
We have made a change of variables from the momenta of $D^{*0}$
and $\bar D^0$ to $\bm{k}$ and the total momentum $\bm{P}$ of the
charm-meson pair.  The dependence on $\bm{P}$  
has been suppressed in Eq.~(\ref{sig-above-thres}).
In the remaining Lorentz-invariant phase-space 
measure $d \phi_{D^* \bar D + {\rm all}}$, $D^* \bar D$
is treated as a single particle with mass 
$M_{D^{*0}} + M_{D^0} \approx M_X$.
According to the Migdal-Watson theorem \cite{Migdal-Watson}, 
the dramatic dependence of the T-matrix element in 
Eq.~(\ref{sig-above-thres}) on $\bm{k}$ 
due to final-state interactions resides in a 
multiplicative factor of the elastic scattering amplitude
$f(k^2/2 \mu)$ given by Eq.~(\ref{f-E}).
A factorization formula can be derived from Eq.~(\ref{sig-above-thres}) 
by dividing the T-matrix element by $f$
and then multiplying the right side by a compensating factor of
$|f|^2$. Since ${\cal T}/f$ is insensitive 
to the small relative momentum $\bm{k}$, we can take the limit 
$\bm{k} \to 0$ in that factor.  The resulting factorization formula is
\beq
d \sigma[D^{*0} \bar D^0(\bm{k}) ] =
\frac{1}{\rm flux} \sum_{\rm all} \int d \phi_{D^* \bar D + {\rm all}} 
\left| {\cal T}[D^{*0} \bar D^0(0) \,+\, {\rm all}]/f(0) \right|^2  
\times \frac{1}{k^2 + 2 \mu E_X}
\frac{d^3k}{(2 \pi)^3 2 \mu}.
\label{dsigDD-fact}
\eeq
The factorization formula for $D^0 \bar D^{*0}$ is given by the 
identical expression.  
These factorization formulas hold for $k < \Lambda$, where $\Lambda$ 
is the momentum scale set by the range of the interaction between the charm mesons.
They indicate that the cross section integrated over $k$ up to 
$k_{\rm max}$ does not have the naive scaling behavior $k_{\rm max}^3$ 
expected from phase space.  Instead it scales like $k_{\rm max}$
in the range $\sqrt{2 \mu E_X} < k_{\rm max} < \Lambda$.

In the corresponding factorization formula for the 
production of $X(3872)$, 
the short-distance factor is the same as in Eq.~(\ref{dsigDD-fact}).
The long-distance factor can be deduced by using the fact that the 
spectral function for the resonance is proportional to the 
imaginary part of the scattering amplitude in Eq.~(\ref{f-E}):
\beq
{\rm Im} f(E) = \frac{\pi}{\mu a} \delta( E + 1/(2 \mu a^2))
+ \frac{\sqrt{2 \mu E}}{1/a^2 + 2 \mu E} \theta(E) .
\label{Imf-E}
\eeq
The term with the factor $\theta(E)$ is the combined contribution from 
$D^{*0} \bar D^0$ and $D^0 \bar D^{*0}$.  The term with the delta function
is the contribution from $X(3872)$.  The long-distance factor 
in the factorization formula for $D^{*0} \bar D^0$ in 
Eq.~(\ref{dsigDD-fact}) is the $\theta(E)$ term in Eq.~(\ref{Imf-E})
multiplied by $dE/(4 \pi^2)$. 
In the corresponding factorization formula for $X(3872)$,
the long-distance factor is the $\delta$-function term 
in Eq.~(\ref{Imf-E}) multiplied by $dE/(2 \pi^2)$ and integrated over $E$.
The resulting factorization formula is
\beq
\sigma[ X(3872) ] =
\frac{1}{\rm flux} \sum_{\rm all} \int d \phi_{D^* \bar D + {\rm all}} 
\left| {\cal T}[D^{*0} \bar D^0(0) \,+\, {\rm all}]/f(0) \right|^2  
\times \frac{\sqrt{2 \mu E_X}}{2 \pi \mu}.
\label{sigX-fact}
\eeq
The dependence on the momentum $\bm{P}$ of the $X(3872)$ has been
suppressed, but $\bm{P}$ can be identified with the total momentum 
of $D^{*0} \bar D^0$ in the short-distance factor.

Since the long-distance factor in Eq.~(\ref{sigX-fact}) is
proportional to $E_X^{1/2}$,
the cross section goes to $0$ as the binding energy goes to $0$. 
This is in accord with the common intuition that the production 
of a weakly-bound molecule should be suppressed in high-energy
collisions. However the degree of suppression is often overestimated.
The common intuition, as exemplified by Ref.~\cite{Bignamini:2009sk}, 
is that the suppression factor should scale like $E_X^{3/2}$. 
In the case of an $S$-wave threshold resonance,
there is a much milder suppression factor proportional to $E_X^{1/2}$.

A factorization formula equivalent to that in Eq.~(\ref{sigX-fact})
was previously derived in Ref.~\cite{Braaten:2005jj}.
In the factorization formula in Ref.~\cite{Braaten:2005jj},
the long-distance factor is also proportional to $E_X^{1/2}$.
However the expression for the short-distance factor 
includes an explicit factor of $\Lambda^2$, where $\Lambda$ is the 
ultraviolet cutoff of an effective field theory for
charm mesons with a large scattering length.  That dependence 
on $\Lambda$ is presumably cancelled by other terms 
in the short-distance factor that depend implicitly on $\Lambda$.
Thus the expression for the short-distance factor
in Ref.~\cite{Braaten:2005jj} can not be interpreted literally.
The advantage of the factorization formula in Eq.~(\ref{sigX-fact})
is that the short-distance factor is well-defined.

\section{Production at the Tevatron}
\label{sec:CDFdata}

In this section, we use results from the CDF Collaboration
to estimate the cross sections for $X(3872)$
in $p \bar p$ collisions at the Tevatron from both prompt 
QCD mechanisms and from decays of $b$ hadrons.

Within a few months of the discovery of the $X(3872)$ 
by the Belle Collaboration~\cite{Choi:2003ue}, its existence 
was confirmed by the CDF Collaboration through inclusive production 
in $p \bar p$ collisions at the Tevatron~\cite{Acosta:2003zx}.
The $X(3872)$ was also observed at the Tevatron by the 
$D\emptyset$ Collaboration~\cite{Abazov:2004kp},
who showed that many of its production characteristics 
are similar to those of the $\psi(2S)$.
The CDF Collaboration subsequently showed that for a 
specific data sample consisting of $X(3872)$
in the decay channel $J/\psi \pi^+ \pi^-$ with a modest $p_T$ cut, 
the production is dominated by prompt
QCD mechanisms rather than $b$-hadron decays~\cite{Bauer:2004bc}.
The information given about this data sample can be used to 
estimate the cross section.

In Ref.~\cite{Bauer:2004bc}, the CDF collaboration observed both 
the $X(3872)$ and the $\psi(2S)$  in a $J/\psi \pi^+ \pi^-$ data sample 
with a cut $p_T(J/\psi)>4$~GeV as well as cuts on the pion momenta. 
For the $X(3872)$, they applied an additional cut 
$M(\pi^+ \pi^-)>0.5$~GeV to increase the signal-to-background ratio.
For both the $X(3872)$ and $\psi(2S)$, they measured the long-lived fraction
$f_{\rm LL}$ that come from the decay of $b$ hadrons: 
$f_{LL}^X=16.1 \pm 4.9 \pm 2.0 \%$
and $f_{\rm LL}^{\psi}=28.3 \pm 1.0 \pm 0.7 \%$.
The complementary fractions $1-f_{\rm LL}$ are produced promptly 
by QCD mechanisms.  Using the numbers of events reported in 
Ref.~\cite{Bauer:2004bc}, we determine the product of the    
cross section for $X(3872)$ and its branching fraction
into $J/\psi \pi^+\pi^-$  for both the prompt and $b$-decay mechanisms:                                                         
\begin{subequations}
\begin{eqnarray}
\sigma_{\rm prompt} [X(3872) ]~{\rm Br} [X \rightarrow J/\psi \pi^+ \pi^- ] 
& = & (0.0335 \pm 0.0055) (\epsilon_\psi/\epsilon_X) \, \sigma_{\rm total}[\psi(2S)],
\label{exp_sigma_X-1} \\
\sigma_{b \textrm{-decay}} [X(3872) ]~{\rm Br} [X \rightarrow J/\psi \pi^+ \pi^- ]
&=& (0.00643 \pm 0.0023) (\epsilon_\psi/\epsilon_X) \, \sigma_{\rm total}[\psi(2S)],
\label{exp_sigma_b_X-1}
\end{eqnarray}
\label{exp_sigma-1}
\end{subequations}
where $\epsilon_X$ and $\epsilon_\psi$ are the efficiencies for observing 
$X(3872)$ and $\psi(2S)$ events in this decay sample. 
The ratio $\epsilon_\psi/\epsilon_X$ is likely to deviate from 1 
by tens of percents rather than factors of 2~\cite{Bauer:2004bc}.
The ``total'' cross section for $\psi(2S)$ in Eqs.~(\ref{exp_sigma-1})
is the sum of the prompt and $b$-decay cross sections.
The cross sections in Eqs.~(\ref{exp_sigma-1}) should be interpreted 
as those for production of $X(3872)$ and $\psi(2S)$ within the cuts 
used to define the two data samples.  The cut $p_T(J/\psi)>4$~GeV 
implies constraints on the transverse momenta of the $X(3872)$
and $\psi(2S)$.  The decay of $X(3872)$ into  $J/\psi \pi^+ \pi^-$ 
proceeds through its decay into $J/\psi \rho^*$, where $\rho^*$ is a 
virtual $\rho$ meson, with $J/\psi$ nearly at rest in the $X$ rest frame. 
Thus the momentum components of $X$ and $J/\psi$ differ only by the factor 
$M_{X}/M_{J/\psi}$.  We can therefore interpret the cross section on the 
left side of Eqs.~(\ref{exp_sigma_X-1}) as the prompt cross section 
for $X \rightarrow J/\psi \pi^+ \pi^-$ with $p_T(X)>5.0$~GeV.
In the decay $\psi(2S) \to J/\psi \pi^+ \pi^-$, the momentum of the $J/\psi$ 
in the $\psi(2S)$ rest frame can range up to $0.5$~GeV. 
Using Monte Carlo simulations, we have checked that the cross section for 
$\psi(2S) \to J/\psi \pi^+ \pi^-$ with $p_T(J/\psi)>4$~GeV can be approximated 
by the cross section with $p_T(\psi(2S))>5$~GeV within $12\%$ accuracy.

To obtain estimates of the cross sections for $X(3872)$ from
Eqs.~(\ref{exp_sigma-1}), we need a measurement of the cross section 
for $\psi(2S)$.  The sum of the prompt and $b$-decay cross sections 
for $\psi(2S)$ in the phase-space region 
$p_T(\psi(2S))>5$~GeV and $|y(\psi(2S))|<0.6$
has been measured by the CDF Collaboration \cite{Aaltonen:2009dm}.
The product of this cross section and the branching fraction 
for $\psi(2S)$ into $\mu^+ \mu^-$ is $0.69 \pm 0.01 \pm 0.06$~nb.
The branching fraction into muons is 
$(7.5 \pm 0.8) \times 10^{-3}$~\cite{Amsler:2008zzb}.
If we assume that the rapidity distribution for the $X$ and the 
$\psi(2S)$ are similar in the acceptance region of the CDF detector, 
we can insert this cross section measurement
for $\psi(2S)$  into Eqs.~(\ref{exp_sigma-1}). Using the estimate 
$\epsilon_\psi/\epsilon_X \approx 1$, we find
\begin{subequations}
\begin{eqnarray}
\sigma_{\rm prompt}[X(3872)]~{\rm Br}[X\rightarrow J/\psi \pi^+ \pi^-] 
& \approx  & 3.1 \pm 0.7~{\rm nb},
\label{exp_sigma_X-2} \\
\sigma_{b \textrm{-decay}} [X(3872) ]~{\rm Br} [X \rightarrow J/\psi \pi^+ \pi^- ]
   & \approx &
0.59 \pm 0.23 \, {\rm nb}.
\label{exp_sigma_b_X-2}
\end{eqnarray}
\label{exp_sigma-2}
\end{subequations}
We interpret these as the cross sections for 
$X(3872) \rightarrow J/\psi \pi^+ \pi^-$
with $p_T(X)>5$~GeV and $|y(X)|<0.6$. 
The error bars come only from the uncertainties in the $\psi(2S)$ 
cross sections and the numerical factors in Eqs.~(\ref{exp_sigma-1}).
They do not include the systematic error
from setting $\epsilon_\psi/\epsilon_X=1$, which may be tens of percent.
They also do not include the systematic error from approximating 
the cross section for $\psi(2S) \to J/\psi \pi^+ \pi^-$
with $p_T(J/\psi) > 4$~GeV 
by the cross section with $p_T(\psi(2S)) > 5$~GeV.
The estimate for the prompt cross section in Eq.~(\ref{exp_sigma_X-2}) 
agrees with the estimate given in Ref.~\cite{Bignamini:2009sk}.  

Existing measurements provide some constraints on the 
branching fraction for the $X(3872)$ into $J/\psi \pi^+ \pi^-$. 
The Babar Collaboration has set a lower bound 
${\rm Br} >0.042$ at $90\%$ C.L.~\cite{Aubert:2005vi}. 
The sum of the measured branching 
ratios for $J/\psi \pi^+ \pi^- \pi^0$~\cite{Abe:2005ix}, 
$J/\psi \gamma$~\cite{Abe:2005ix,:2008rn}, 
$D^0 \bar D^0 \pi^0$~\cite{Aubert:2008gu,Belle:2008te} and 
$\psi(2S) \gamma$~\cite{:2008rn} relative to 
$J/\psi \pi^+ \pi^-$ is $13.5\pm2.9$. If taken at face value, 
this implies an upper bound on the branching fraction into 
$J/\psi \pi^+ \pi^-$: ${\rm Br} <0.093$ at $90\%$ C.L. 
If we allow  ${\rm Br}[X\rightarrow J/\psi \pi^+ \pi^-]$ 
to range from $0.042$ to $0.093$, our estimate of the prompt cross section 
for $X(3872)$ from Eq.~(\ref{exp_sigma_X-2}) ranges from 72~nb to 33~nb,
while our estimate of the $b$-decay cross section 
from Eq.~(\ref{exp_sigma_b_X-2}) ranges from 14~nb to 6~nb.

\section{Charm-meson-pair cross Sections and the $\bm{X(3872)}$}
\label{sec:charmpair}

In this section, we rederive the upper bound on the cross section 
for producing a charm-meson molecule presented in 
Ref.~\cite{Bignamini:2009sk}.  We point out that their calculation 
of the upper bound for the $X(3872)$ is too small by orders of magnitude, 
because they do not take into acount charm-meson rescattering.
We also present an alternative method for calculating the upper bound 
that is much more efficient than the method 
used in Ref.~\cite{Bignamini:2009sk}.
Finally, we derive an order-of-magnitude estimate of the  
cross section for producing $X(3872)$ and show that 
this estimate is compatible with the
observed prompt production rate at the Tevatron.

\subsection{Upper Bound on Prompt Cross Section}
\label{sec:upper}

In Ref.~\cite{Bignamini:2009sk}, Bignamini et al.\  derived 
an upper bound on the cross section for producing a 
charm-meson molecule in terms of the
cross section for producing charm-meson pairs
integrated over a region of small relative momentum.
In the case of prompt production of $X(3872)$ at the Tevatron,
their upper bound was lower than the observed production rate
by orders of magnitude, 
casting doubt on the identification of the $X(3872)$ as a 
loosely-bound $D^{*0}\bar D^{0}$ molecule. 

The starting point in the derivation of the upper bound in 
Ref.~\cite{Bignamini:2009sk} was an expression for the 
inclusive cross section for the production of $X(3872)$ in
terms of its momentum-space wavefunction
$\tilde \psi_X(\bm{k})$:
\beq
\sigma[X ] = 
\frac{1}{\rm flux} \sum_{\rm all} \int d\phi_{X+{\rm all}} 
\left| \int^{k_{\rm max}} \frac{d^3k}{(2 \pi)^3 \sqrt{2\mu}} \,
      {\cal T}[(D^{*0} \bar D^0)_+(\bm{k}) \, + \, {\rm all} ] \,
      \tilde \psi_X(\bm{k}) \right|^2 .
\label{sig-0}
\eeq
The notation $(D^{*0} \bar D^0)_+$ indicates a projection of the 
charm mesons onto the even-charge-conjugation component defined
by the right side of Eq.~(\ref{X-DD}).
The integral over the relative momentum $\bm{k}$ has been 
restricted to the region $|\bm{k}| < k_{\rm max}$ 
in which the integrand has significant support.
By applying the Schwartz inequality to the integral
over $\bm{k}$ in Eq.~(\ref{sig-0}), one obtains an upper bound
\bea
\sigma[X] &\leq& \frac{1}{\rm flux} \sum_{\rm all} \int d \phi_{X+{\rm all}}  
\int^{k_{\rm max}} \frac{d^3k}{(2 \pi)^3 2\mu} \,  
\left| {\cal T}[(D^{*0} \bar D^0)_+(\bm{k}) \,+\, {\rm all}] \right|^2 \,
\nonumber
\\
&& \times \, \int^{k_{\rm max}} \frac{d^3k}{(2 \pi)^3} \, 
\left| \tilde{\psi}_X(\bm{k}) \right|^2 .
\label{upper_bound}
\eea
The last factor is less than 1, because it is simply the incomplete 
normalization integral for the wavefunction. 
The remaining expression on the right side can be expressed as the sum
of inclusive cross sections for producing pairs of charm mesons
plus interference terms: 
\bea
&& \frac{1}{{\rm flux}} \sum_{\rm all} \int d \phi_{X+{\rm all}}  \int^{k_{\rm max}}
\frac{d^3k}{(2 \pi)^3 2\mu} \,  
\left| {\cal T}[(D^{*0} \bar D^0)_+(\bm{k}) \,+\, {\rm all}] \right|^2 \,
\nonumber
\\
&& \approx \mbox{$\frac12$} \sigma[D^{*0} \bar D^0(k<k_{\rm max})]
       + \mbox{$\frac12$} \sigma[D^{0} \bar D^{*0}(k<k_{\rm max})] 
       + ({\rm interference})\, .
\label{charge_conj_state}
\eea
In a high energy collision, an inclusive cross section is summed 
over many additional particles, which we have represented by ``all''. 
The interference terms between the T-matrix elements 
for $D^{*0} \bar D^0$ and $D^{0} \bar D^{*0}$ should average to $0$
upon summing over all these additional particles. 
Thus the upper bound in 
Eq.~(\ref{upper_bound}) can be written in a very simple form:
\beq
\sigma[X] \leq
\mbox{$\frac12$} \sigma[D^{*0} \bar D^0(k < k_{\rm max})]
+\mbox{$\frac12$} \sigma[D^0 \bar D^{*0}(k < k_{\rm max})].
\label{sig-1}
\eeq
This inequality 
holds provided $k_{\rm max}$ is large enough to provide
most of the region of support for the integral in Eq.~(\ref{sig-0}).
In the case of $p \bar p$ collisions, charge conjugation symmetry
implies that the two cross sections on the right side of 
Eq.~(\ref{sig-1}) are equal.

In Ref.~\cite{Bignamini:2009sk}, the authors assumed that the
region of support for the integral over $\bm{k}$ in 
Eq.~(\ref{sig-0}) extends only up to the momentum scale 
set by the binding momentum $(2 \mu E_X)^{1/2}$
of the molecule.  The central value of the binding energy 
from Eq.~(\ref{MX}), $E_X = 0.30$~MeV, corresponds to a
binding momentum of 24~MeV. 
The authors of Ref.~\cite{Bignamini:2009sk} took the upper 
endpoint of the region of support for the 
integral in Eq.~(\ref{sig-0}) to be $k_{\rm max} = 35$~MeV.
They used the Monte Carlo event generators Pythia and Herwig
to calculate the distribution in the relative momentum $k$ 
for prompt charm-meson pairs $D^{*0} \bar D^0$ at the Tevatron.
The  distribution was normalized using CDF data 
on the production of $D^0 D^{*-}$ at the 
Tevatron~\cite{Reisert:2007zza}. 
For the charm-meson-pair cross section integrated up to 
$k_{\rm max} = 35$~MeV, they obtained $0.11$~nb and $0.071$~nb 
using Pythia and Herwig, respectively.
These cross sections are more than two orders of magnitude smaller than 
our estimate of the prompt cross section for $X(3872)$ from 
Eq.~(\ref{exp_sigma_X-2}), which is $33-72$~nb.
They concluded that the $X(3872)$ is unlikely to be a charm-meson
molecule.

The flaw in the argument of Ref.~\cite{Bignamini:2009sk}
is that the authors ignored the effects of rescattering
of the charm-meson pair.
Rescattering can be important for relative momenta 
comparable to or smaller than the scale $\Lambda$ set by the 
range of the interaction.
For a generic molecular state whose binding momentum 
is of the order $\Lambda$, rescattering does not change the momentum scale.
However an S-wave threshold resonance is characterized by a very
large scattering length $a$ that arises from a fine-tuned balance 
between the interaction strength of the constituents 
and the effects of rescattering.  The rescattering effects are so 
strong that the cross section for elastic scattering saturates 
the unitarity bound for relative momentum in the range between
$(2 \mu E_X)^{1/2}$ and $\Lambda$.
These same effects allow charm mesons that are created with 
relative momenta as large as $\Lambda$ to rescatter into 
small relative momenta of order $\sqrt{2\mu E_X }$.
For this reason, the region of support for the integral over the 
relative momentum in Eq.~(\ref{sig-0}) extends up to the
scale $\Lambda$. This dramatically increases the upper bound on the 
cross section given by Eq.~(\ref{sig-1}).
Now the charm-meson-pair cross sections 
calculated using Monte Carlo methods  in 
Ref.~\cite{Bignamini:2009sk} scale like $k_{\rm max}^3$ 
from the phase space of the charm-meson pair.  
If we take $m_\pi$ as an estimate of the scale $\Lambda$
and replace the value 
$k_{\rm max} = 35$~MeV used in Ref.~\cite{Bignamini:2009sk} 
by $k_{\rm max} = c \Lambda$, the charm-meson-pair 
cross section of $0.07-0.11$~nb is increased by a factor of 
$64 c^3$.  For reasonable choices of $c$,
this upper bound can be larger than the estimated prompt cross
section of $33-72$~nb obtained in Section~\ref{sec:CDFdata}.

There is an alternative way to see that $k_{\rm max}$ in the upper 
bound in Eq.~(\ref{sig-1}) must be much larger 
than the binding momentum of the $X(3872)$.
By the Migdal-Watson theorem \cite{Migdal-Watson}, 
the dramatic dependence of the T-matrix element in 
Eq.~(\ref{sig-0}) on $\bm{k}$ resides in a multiplicative factor 
of the elastic scattering amplitude $f(k^2/2 \mu)$.
If the region of support of the integral over $\bm{k}$ 
in Eq.~(\ref{sig-0}) was limited to the momentum scale
$(2 \mu E_X)^{1/2}$, we could replace $f(k^2/2 \mu)$
by the universal scattering amplitude in Eq.~(\ref{f-E})
and $\tilde \psi_X(\bm{k})$ by the universal momentum-space 
wavefunction in Eq.~(\ref{psi-uni:k}).  However the resulting 
integral over $\bm{k}$ in Eq.~(\ref{sig-0}) 
is logarithmically ultraviolet divergent.
This is a signal that the region of support for this integral 
is not limited to $k$ of order $\sqrt{2 \mu E_X}$.
It extends up to the momentum scale set by the range of 
interactions between the charm mesons.

\subsection{More efficient calculation of upper bound}
\label{sec:upperbound}

The upper bound on the cross section for the $X(3872)$
in Eq.~(\ref{sig-1}) is the cross section for charm-meson pairs
integrated over the relative momentum $k$ up to $k_{\rm max}$. 
In Ref.~\cite{Bignamini:2009sk}, that upper bound was calculated 
for the case of prompt production of $X(3872)$ at the Tevatron.
We now present an alternative calculation of that charm-meson-pair 
cross section using a method that is much more efficient.
In the analysis of Ref.~\cite{Bignamini:2009sk}, 
the momentum distribution for charm-meson pairs 
was calculated by generating 
more than $5 \times 10^{10}$ 2-to-2 parton-level events
and then passing them through Pythia~\cite{Sjostrand:2006za}
or Herwig~\cite{Corcella:2000bw} for showering and hadronization.
Only a tiny fraction of these events include a pair of  
nearly-collinear charm mesons with substantial $p_T$. 
These specific events are generated by an outgoing 
gluon from the 2-to-2 parton 
collision fragmenting into a charm-quark pair in the showering process, 
which in turn hadronizes most of the time into a pair of charm mesons.
We calculate the charm-meson-pair cross section by generating 
$g g \rightarrow g c \bar c$ parton-level events with 
MadGaph~\cite{Alwall:2007st}
and then passing them through Pythia for showering and hadronization. 
This 2-to-3 parton-level process is the dominant
mechanism for the production of a pair of
charm mesons with small relative momentum and substantial $p_T$.
In order to further increase the efficiency, only parton-level events 
for which the $c \bar c$ pair have relative momentum 
below 2~GeV have been generated. 
We checked that this reduction of the parton-level phase space
does not affect the hadron-level cross section 
for charm mesons with small relative momentum $k<600$ MeV.
The main advantage of generating only $g g \rightarrow g c \bar c$
events is that the momentum 
distribution in the region of small relative momentum can be calculated 
with accuracy comparable to that in Ref.~\cite{Bignamini:2009sk}
by generating fewer events by about a factor of $10^4$.

\begin{figure}[t]
\centerline{\includegraphics*[height=8cm,angle=0,clip=true]{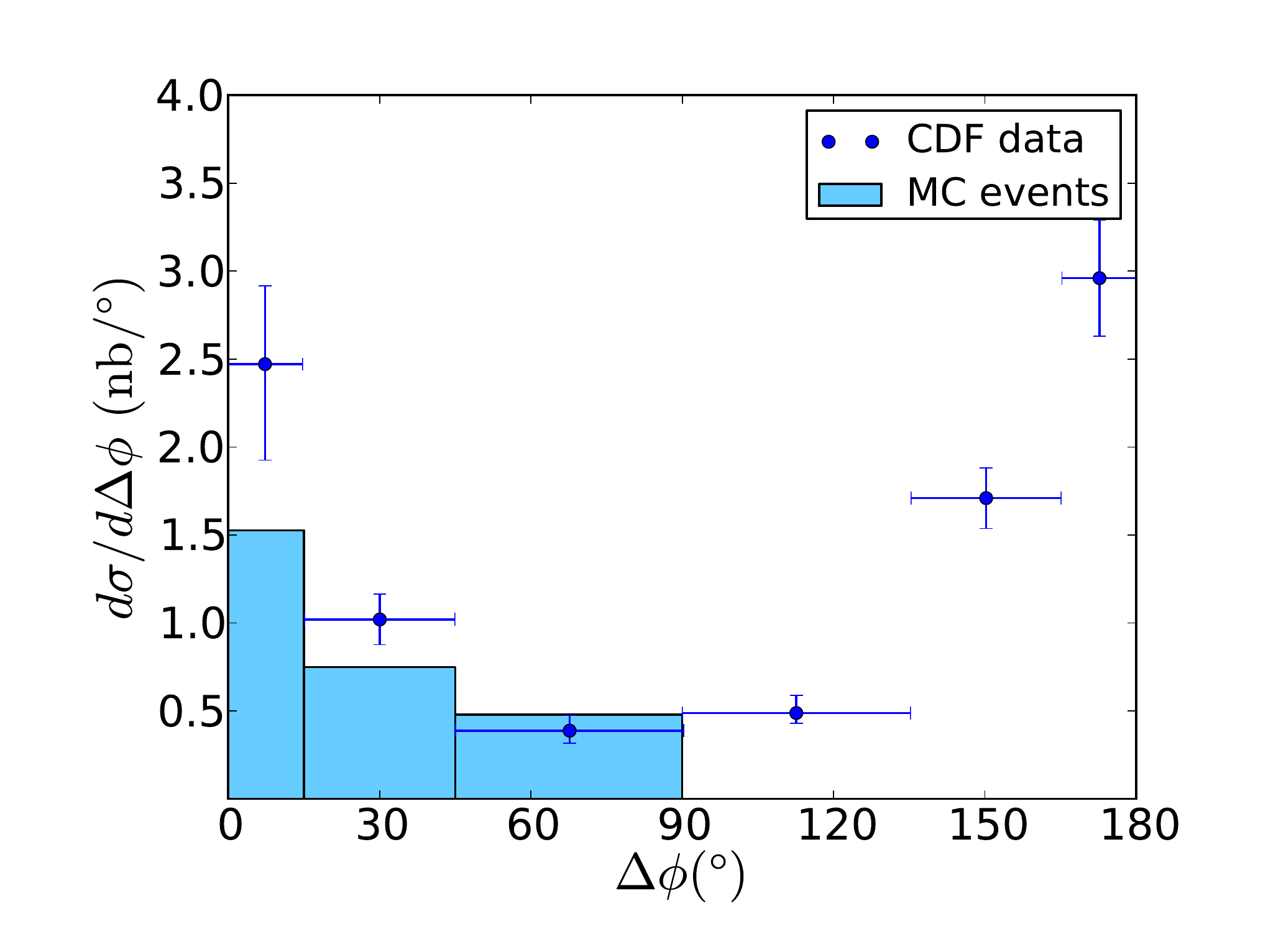}}
\vspace*{0.0cm}
\caption{Cross section for the inclusive production of 
$D^{0}D^{* -} $ in $p \bar p$ collisions at $\sqrt{s} = 1.96$~TeV.
The cross section is integrated over bins of  
$\Delta \phi$ -- the angle between the transverse momenta 
of $D^{0}$ and $D^{* -}$.  The data points are 
measurements by the CDF Collaboration at the 
Tevatron \cite{Reisert:2007zza}. The histogram is calculated using the 
Monte Carlo methods described in the text.}
\label{fig1}
\end{figure}

We follow Ref.~\cite{Bignamini:2009sk} in normalizing the 
Monte Carlo distribution using measurements by the CDF Collaboration 
of the inclusive cross section for $D^{0} D^{* -}$ in $p \bar p$ 
collisions at the Tevatron~\cite{Reisert:2007zza}.
The CDF Collaboration measured this cross section in the phase-space region 
$p_T(D^{0}), \, p_T(D^{*-}) >5.5$~GeV and $|y(D^{0})|, |y(D^{*-})|\, <1$.
The cross section differential in $\Delta \phi$ -- 
the angle between the transverse momenta of the two charm mesons 
-- is shown in Fig.~\ref{fig1}. 
In the analysis of Ref.~\cite{Bignamini:2009sk},
the Monte Carlo distribution  for pairs of charm mesons was 
normalized to this data.  Since the error bars are smallest in the 
bins of $\Delta \phi$ closest to $90\degree$, this means that in practice
the Monte Carlo distribution in Ref.~\cite{Bignamini:2009sk} 
was normalized to the data near $90\degree$.  In our analysis,
as we are primarily interested in configurations in which two charm mesons 
are almost collinear, we normalize the Monte Carlo event rate instead
to the measured rate in the first bin in $\Delta \phi$, which extends from 
$0$ to $15\degree$.  
Compared to normalizing the rate near $90\degree$,
this increases the normalization by about 
$60\%$ at the expense of a $20\%$ error from the experimental uncertainty. 
Our Monte Carlo events for $D^{0} D^{*-}$ were generated 
via the parton-level process $gg \rightarrow g c \bar c$, 
consistent with our procedure for generating $D^{*0} \bar D^{0}$ events. 
In order to increase the efficiency, we applied the parton-level cuts
$p_T(c), \, p_T(\bar c)>3.5$~GeV, $|y(c)|, \, |y(\bar c)|<2$, 
and $\Delta \phi(c, \bar c)<90 \degree$,
which do not affect the $D^{0}D^{* -}$ production rate in the region 
of interest.  The rescaling factor -- defined by the ratio of the 
experimental rate over the Monte Carlo rate in the first bin in 
$\Delta \phi$ -- is $1.62$.  This same rescaling factor was then applied 
to the distributions of $D^{*0} \bar D^0$ events.

\begin{figure}[t]
\includegraphics*[width=8cm,angle=0,clip=true]{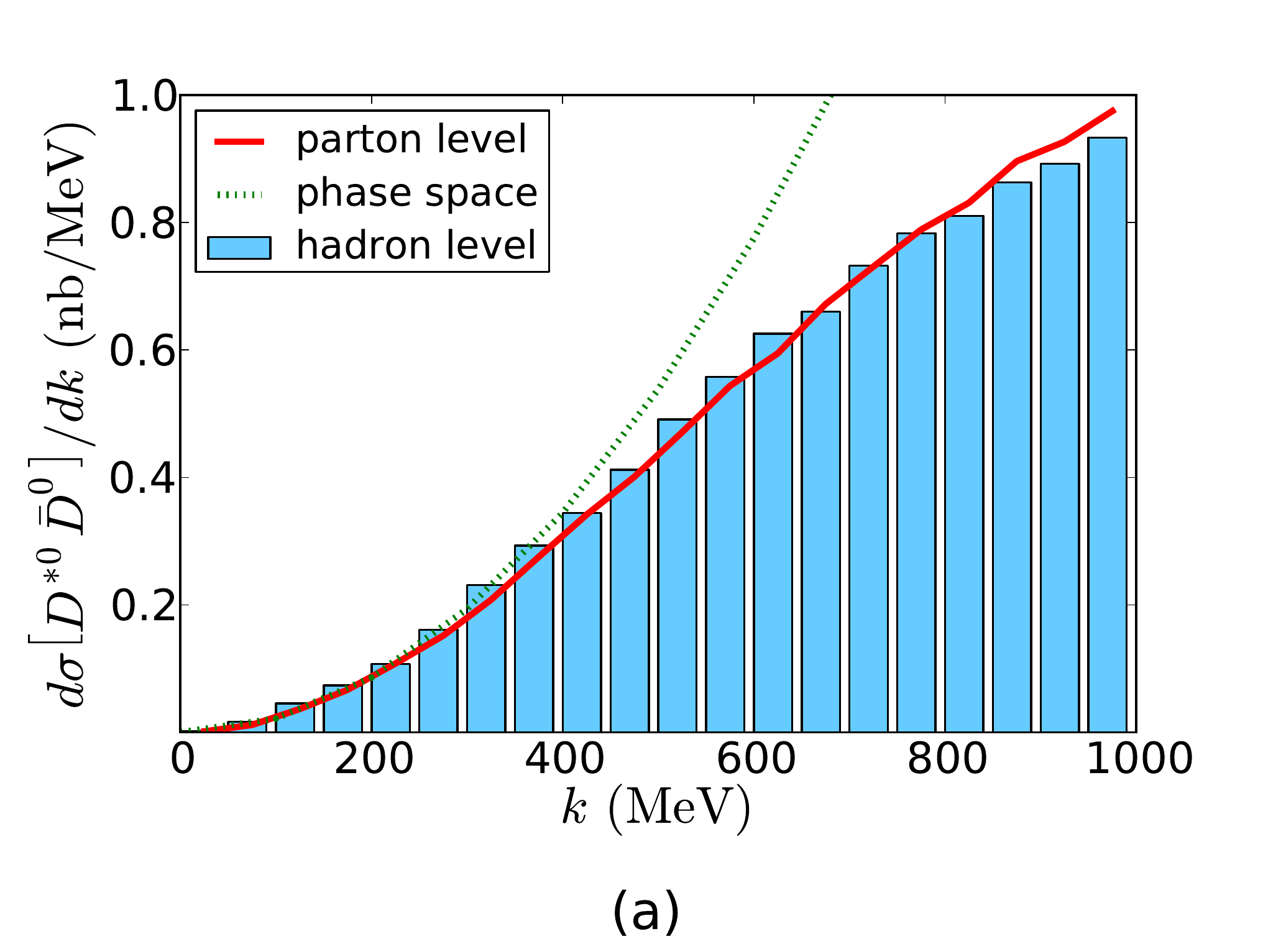}
\includegraphics*[width=8cm,angle=0,clip=true]{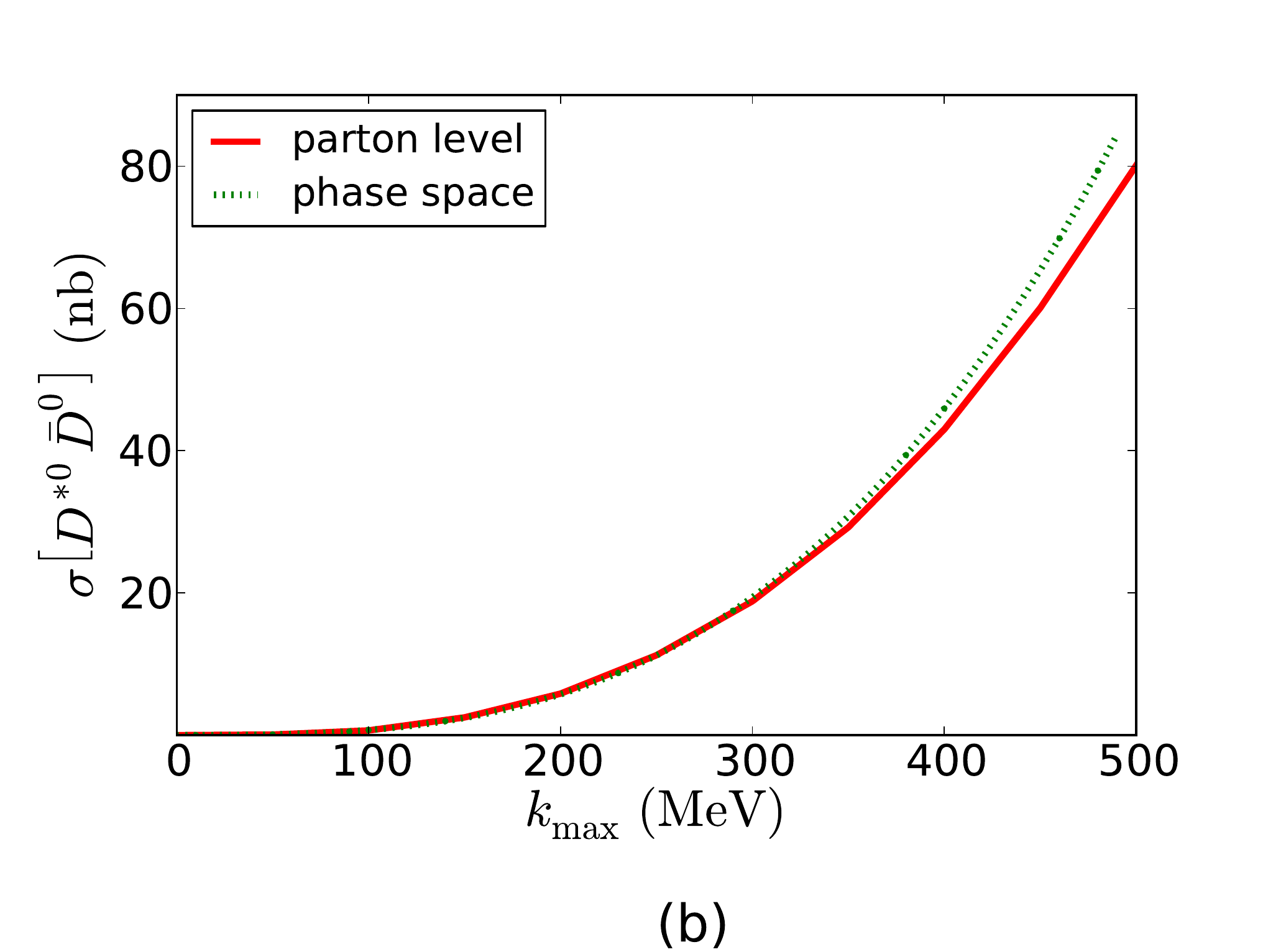}
\vspace*{0.0cm}
\caption{Cross sections for the inclusive production of 
$D^{*0} \bar{D}^0$ in $p \bar p$ collisions at $\sqrt{s} = 1.96$~TeV.
(a) Differential cross section integrated over the region 
$p_T(D^{*0}\bar{D}^0) > 5$~GeV and $|y(D^{*0}  \bar{D}^0)| < 0.6$ 
as a function of the relative momentum $k$ of $D^{*0} \bar{D}^0$. 
The histogram is the hadron-level cross section
calculated using the Monte Carlo method described in the text.
The solid curve is the parton-level differential cross section
integrated over the region 
$p_T(c  \bar c) > 5$~GeV and $|y(c  \bar c)| < 0.6$ 
and normalized to the hadron-level cross section
in the region $k<1$~GeV.
(b) Cross section obtained by integrating the parton-level 
differential cross section up to $k_{\rm max}$.
The dotted curves are simple phase-space distributions.}
\label{fig2}
\end{figure}

Having fixed the normalization of our Monte Carlo distributions, 
we proceed to calculate the production rate of $D^{*0} \bar D^0$
with small relative momentum at the Tevatron, run II. 
The differential cross section is displayed as a histogram 
in the relative momentum $k$ in Fig.~\ref{fig2}(a).
The distribution, which was obtained by generating about 
$5 \times 10^{6}$ events, 
is as smooth as the distribution in Ref.~\cite{Bignamini:2009sk}, 
which was obtained by generating more than 
$5 \times 10^{10}$ events.
For $k$ less than about 300~MeV, the shape of the Monte Carlo distribution 
is consistent with $k^2$, in accord with simple phase-space suppression.

The results of our calculation using MadGraph and the event generator 
Pythia can be reproduced approximately 
using a simpler method that does not require any event generator.
We assume that a $D^{*0} \bar D^0$ pair with small relative momentum $k$
is formed from a $c \bar c$ pair with approximately 
the same relative momentum $k$.
The resulting approximation for the $D^{*0} \bar{D}^0$ cross section 
is proportional to the parton-level cross section for a $c \bar c$ pair:
\beq
\sigma[D^{*0} \bar D^0(k < k_{\rm max}) ] \approx
\sigma[c \bar c(k < k_{\rm max}) ]\ 
\, \times \, F[c \bar c \rightarrow D^{*0} \bar D^{0}]  .
\label{sidDDcc}
\eeq
The factor $F[ c\bar c \rightarrow D^{*0} \bar D^0]$ 
can be interpreted 
as the probability for a nearly-collinear charm-quark pair 
to evolve into a nearly-collinear $D^{*0} \bar D^{0}$ pair. 
The shape of this parton-level distribution agrees well with the 
hadron-level distribution shown as a histogram in Fig.~\ref{fig2}(a).  
The normalization also agrees if we set 
$F[ c\bar c \rightarrow D^{*0} \bar D^0] = 0.142$.
The resulting distribution is shown as a solid line in Fig.~\ref{fig2}(a). 
Alternatively, the normalizing factor can be determined 
from the CDF measurement of the cross section for $D^0 \bar D^{*-}$ 
in the first bin in $\Delta \phi$ in Fig.~\ref{fig1}. 
The resulting value is $F[c \bar c \rightarrow D \bar D^{*}]=0.118$,
which is smaller than the value obtained 
normalizing to the hadron-level distribution by about $17 \%$.

To obtain the charm-meson-pair cross section 
$\sigma[  D^0 \bar D^{*0}(k<k_{\textrm{max}}) ]$ 
that appears in the upper bound in Eq.~(\ref{sig-1}),
the $k$ distribution in Fig.~\ref{fig2}(a) 
must be integrated up to $k_{\textrm{max}}$.
In Fig.~\ref{fig2}(b), we show the cross section 
in Eq.~(\ref{sidDDcc}) with 
$F[ c\bar c \rightarrow D^{*0} \bar D^0] = 0.142$ 
as a function of $k_{\textrm{max}}$. 
For the value $k_{\rm max}=35$~MeV used in Ref.~\cite{Bignamini:2009sk},
we obtain a cross section of $0.030$~nb, which is smaller than 
that obtained in Ref.~\cite{Bignamini:2009sk} using Pythia 
by a factor of 3.7.

\subsection{Estimate of the Prompt Cross Section}

The factorization formula for the production of $X(3872)$ 
in Eq.~(\ref{sigX-fact}) separates the short-distance
effects associated with the production of the charm mesons 
from long-distance effects associated with their binding into 
the $X(3872)$.  The Monte Carlo methods used to calculate the 
upper bound on the prompt cross section at the Tevatron 
can be combined with this factorization formula to obtain an 
order-of-magnitude estimate of the prompt cross section.
The Monte Carlo methods use an event generator, 
such as Pythia or Herwig, to model the hadronization process that
produces the charm mesons.  The tuning of the parameters of the 
event generator implicitly takes into account typical
hadronic cross sections.  It does not take into account 
any unnaturally large cross section, such as that
associated with an S-wave threshold resonance like the $X(3872)$.

The charm-meson pair $D^{*0} \bar D^0$ is expected to have a 
typical hadronic production rate as long as its relative 
momentum $k$ is comparable to or larger than the momentum scale $\Lambda$
set by the range of the interactions between charm mesons.
According to the Migdal-Watson theorem, the production rate 
for $k < \Lambda$ is enhanced because of the factor 
$f(k^2/2 \mu) = (-1/a - i k)^{-1}$ in the T-matrix element.
An event generator like Pythia contains no information about
such an enhancement factor.  When such an event generator is used, 
there is an implicit assumption that the T-matrix element 
${\cal T}[(D^{*0} \bar D^0)_+(\bm{k}) \,+\, {\rm all}]$
does not vary dramatically for $|\bm{k}| < \Lambda$.
If this assumption is applied to Eq.~(\ref{sig-above-thres}),
we find that the ``naive" cross section calculated using an 
event generator and integrated over $|\bm{k}| < \Lambda$ can be
approximated by
\beq
\sigma_{\rm naive}[D^{*0} \bar D^0(k<\Lambda) ] \approx
\frac{1}{\rm flux} \sum_{\rm all} \int d \phi_{D^* \bar D + {\rm all}} 
\left| {\cal T}[D^{*0} \bar D^0(|\bm{k}| = \Lambda) \,+\, {\rm all}] \right|^2 
\times \frac{\Lambda^3}{12 \pi^2 \mu}  ,
\label{sig-naive}
\eeq
where the T-matrix element has been evaluated
at $|\bm{k}| = \Lambda$.

In the short-distance factor of the factorization formula for the 
$X(3872)$ in Eq.~(\ref{sigX-fact}), the term ${\cal T}/f$ 
is evaluated at $\bm{k} = 0$.
However this term is insensitive 
to the relative momentum for  $|\bm{k}| < \Lambda$, so
it can be approximated by its value when $|\bm{k}| = \Lambda$.
The factorization formula can be expressed in the form
\beq
\sigma[ X(3872) ] \approx
\frac{1}{\rm flux} \sum_{\rm all} \int d \phi_{D^* \bar D + {\rm all}} 
\left| {\cal T}[D^{*0} \bar D^0(|\bm{k}| = \Lambda) \,+\, {\rm all}] \, \Lambda \right|^2  
\times \frac{\sqrt{2 \mu E_X}}{2 \pi \mu},
\label{sigX-fact:Lam}
\eeq
where we have used $\Lambda \gg 1/|a|$ to 
approximate $1/f(\Lambda^2/2 \mu)$ by $\Lambda$.
The short-distance factor in Eq.~(\ref{sigX-fact:Lam})
can now be eliminated in favor of the
naive charm meson cross section in Eq.~(\ref{sig-naive}):
\beq
\sigma[ X(3872) ] \approx
\sigma_{\rm naive}[D^{*0} \bar D^0(k<\Lambda) ] 
\times \frac{6 \pi \sqrt{2 \mu E_X}}{\Lambda}.
\label{sigX-fact:naive}
\eeq
If the naive cross section for producing charm mesons has been 
calculated using an event generator,
this expression can be used to estimate the cross section
for producing $X(3872)$.  

In Section~\ref{sec:upperbound}, we used MadGraph and Pythia
to calculate the naive cross section for $D^{*0} \bar D^0$ 
at the Tevatron as a function of $k_{\rm max}$.
Taking the central value $E_X=0.30$~MeV for the binding energy 
from Eq.~(\ref{MX}), our estimate of the 
prompt cross section from Eq.~(\ref{sigX-fact:naive})
is 1.5, 5.9, and 23~nb for $\Lambda = m_{\pi}/2$, $m_{\pi}$, 
and $2 m_{\pi}$, respectively.  The upper end of this range of cross sections
is compatible with the estimate $33-72$~nb
for the prompt cross section at the Tevatron obtained in 
Section~\ref{sec:CDFdata}.

We can use the estimate of the cross section for $X(3872)$ in
Eq.~(\ref{sigX-fact:naive}) to obtain a quantitative estimate 
of the upper endpoint $k_{\rm max}$ of the region of support 
of the integral in Eq.~(\ref{sig-0}).
The upper bound in Eq.~(\ref{sig-1}) can be approximated by the 
naive charm-meson-pair cross section 
$\sigma_{\rm naive}[D^{*0} \bar D^0(k<k_{\rm max}) ]$ provided 
$k_{\rm max} > \Lambda$:  
\beq
\sigma[ X(3872) ] \le
\sigma_{\rm naive}[D^{*0} \bar D^0(k<k_{\rm max}) ] .
\label{sigX-upper}
\eeq
Demanding that the estimate in Eq.~(\ref{sigX-fact:naive}) is less 
than the upper bound in Eq.~(\ref{sigX-upper}) even 
if $\sqrt{2 \mu E_X}$ is as large as $\Lambda$, 
we obtain the condition
$k_{\rm max} > ( 6 \pi)^{1/3} \Lambda = 2.7~\Lambda$.
The naive cross section in Eq.~(\ref{sigX-upper}) is shown 
as a function of $k_{\rm max}$ in Fig.~\ref{fig2}.
Setting $k_{\rm max} = 2.7~\Lambda$, the upper bound on the 
cross section for $X(3872)$ is 4.2, 32, and 230~nb 
for $\Lambda = m_{\pi}/2$, $m_{\pi}$, and $2 m_{\pi}$, respectively.  
This can be much larger than the prompt cross section
observed at the Tevatron.

\section{NRQCD Factorization}
\label{sec:NRQCD}

In this section, we discuss the NRQCD factorization formula
for the inclusive production of $X(3872)$, which 
expresses the cross section as a sum of parton cross sections 
multiplied by phenomenological constants.
We use simplifying assumptions to reduce the phenonomenological 
constants to a single constant that is determined from our 
estimate of the prompt cross section at the Tevatron.
We then use the NRQCD factorization formula 
to predict the differential cross section 
for prompt production of $X(3872)$ at the Tevatron.
We also predict the differential cross section 
for production of $X(3872)$ from $b$-hadron decays at the Tevatron.

\subsection{NRQCD Factorization Formula}

Inclusive production rates for the $X(3872)$ 
satisfy an NRQCD factorization formula, 
which separates the momentum scales of order $m_c$ 
and larger that are involved in the creation of the $c \bar c$ pair 
from all the smaller momentum scales of QCD \cite{Braaten:2004jg}.
The smaller momentum scales include the scale $m_\pi$ associated 
with the formation of the charm mesons and the scale 
$\sqrt{2 \mu E_X}$ associated with their binding into the $X(3872)$.
The NRQCD factorization formula 
has the form \cite{Bodwin:1994jh}
\begin{eqnarray}
\sigma [ X(3872)] = \sum_n \hat \sigma [ c \bar c_n] 
\; \langle {\cal O}_n^X \rangle ,
\label{ff:H}
\end{eqnarray}
where the sum over $n$ extends over the color and angular-momentum 
channels of a $c \bar c$ pair.
The short-distance factors $\hat \sigma$ are inclusive 
cross sections for producing a $c \bar c$ pair with negligible
relative momentum in the channel $n$ together with hard partons.
They can be calculated using perturbative QCD.
The dependence on the momentum $\bm{P}$ of the $X(3872)$
has been suppressed in Eq.~(\ref{ff:H}), 
but $\bm{P}$ can be identified with the total momentum 
of the $c \bar c$ pair.  The long-distance factor 
$\langle {\cal O}_n^H \rangle$ is proportional to the probability for
the $c \bar c$ pair in the channel $n$ to evolve into the 
$c \bar c$ meson $H$ plus soft partons or hadrons. 
It is called an {\it NRQCD matrix element},
because it can be expressed as a matrix element 
of a local operator in an effective field theory for the 
$c \bar c$ sector of QCD called nonrelativistic QCD (NRQCD).
The NRQCD matrix elements can be treated as phenomenological constants.
They are universal, so once they have been determined by
fitting experimental results, the NRQCD factorization formula 
can be used to predict the production rate in other experiments.

The NRQCD factorization formula was originally developed for 
heavy quarkonium \cite{Bodwin:1994jh}.
The relative sizes of the NRQCD matrix elements depend 
on the angular momentum quantum numbers of the quarkonium state,
scaling as specific powers of the typical relative velocity of 
the heavy quark pair.
The pattern of suppression is summarized by a rather intricate set 
of velocity-scaling rules.  
The NRQCD factorization formula can also be applied to pairs 
of charm mesons with small relative momentum \cite{Braaten:2004jg}.
In this case, there should be a hierarchy in the sizes of the
NRQCD matrix elements according to their orbital angular momentum.
The most important matrix elements should be the S-wave matrix
elements. There are four independent S-wave matrix elements:
$\langle {\cal O}_1^X ({}^1S_0) \rangle$,
$\langle {\cal O}_1^X ({}^3S_1) \rangle$,
$\langle {\cal O}_8^X ({}^1S_0) \rangle$, 
and $\langle {\cal O}_8^X ({}^3S_1) \rangle$.
The subscript 1 or 8 on the operator indicates the color channel: 
color-singlet or color-octet.
The argument is the spectroscopic notation $^{2S+1}L_J$
for the angular-momentum channel.  Explicit expressions for these 
operators in terms of NRQCD fields 
are given in Ref.~\cite{Bodwin:1994jh}.
The S-wave matrix elements can be interpreted as probability
densities for a $c \bar c$ pair created at a point in the
specified state to form the hadron $X$ multiplied 
by spin and color factors. For the four NRQCD matrix elements  
listed above, the products of the spin and color factors are 
$2N_c=6, \,3(2N_c)=18, \, N_c^2-1=8$ and $3(N_c^2-1)=24$, respectively.
The large-scattering-length factorization formulas
in Eqs.~(\ref{dsigDD-fact}) and (\ref{sigX-fact}) imply that the relative
sizes of the NRQCD matrix elements for $X(3872)$ should be the same 
as for $D^{*0} \bar D^0$.  Truncating the NRQCD factorization formula 
to the S-wave terms, we get
\begin{eqnarray}
\sigma [ X(3872) ] &\approx& 
\hat \sigma [ c \bar c_1 ({}^1S_0) ]  \;
	\langle {\cal O}_1^X ({}^1S_0) \rangle 
+ \hat \sigma [ c \bar c_1 ({}^3S_1 ) ]  \;
	\langle {\cal O}_1^X ({}^3S_1) \rangle 
\nonumber
\\
&&
+ \hat \sigma [ c \bar c_8 ({}^1S_0)]  \;
	\langle {\cal O}_8^X ({}^1S_0) \rangle 
+ \hat \sigma [ c \bar c_8 ({}^3S_1 )]  \;
	\langle {\cal O}_8^X ({}^3S_1) \rangle .
\label{fft:X2}
\end{eqnarray}

A possible binding mechanism for the $X(3872)$ is the accidental
fine-tuning of the depth of the interaction potential between 
the charm mesons $D^{*0}$ and $\bar D^0$ so that there is a bound state 
very near the threshold. 
Since the mesons $D^{*0}$ and $\bar D^0$ are color-singlets,
the $c \bar c$ pair in the  $D^* \bar D$ system 
has equal probabilities 1/9 of being in a color-singlet state
or in any of the 8 color-octet states.  
It is therefore reasonable to assume that
the probabilities for formation of the $X(3872)$
from a color-singlet $c \bar c$ pair and from any of the 
color-octet $c \bar c$ pairs are equal.
Given the normalizations of the NRQCD operators,
this assumption translates into a ratio of 3/4 between color-singlet 
and color-octet NRQCD matrix elements: 
\begin{subequations}
\begin{eqnarray}
\langle {\cal O}_1^X ({}^3S_1) \rangle  &\approx&  
\frac34 \langle {\cal O}_8^X ({}^3S_1) \rangle ,
\label{O8-O1:triplet}
\\
\langle {\cal O}_1^X ({}^1S_0) \rangle  &\approx&  
\frac34 \langle {\cal O}_8^X ({}^1S_0)  \rangle.
\label{O8-O1:singlet}
\end{eqnarray}
\label{O8-O1}
\end{subequations}

We now consider the dependence of the NRQCD matrix elements on the 
spin quantum number, which can be spin-triplet or spin-singlet. 
In the hadronization stage of the formation of the $X(3872)$, 
a $c \bar c$ pair evolves into the charm-meson pair
$(D^{*0} \bar D^0)_+$ with even charge conjugation defined in 
Eq.~(\ref{X-DD}).  Voloshin has pointed out that the $c \bar c$ pair 
in $(D^{*0} \bar D^0)_+$ is necessarily 
in a spin-triplet state \cite{Voloshin:2004mh}.
The approximate heavy-quark spin symmetry of QCD implies that 
transitions that change the spin state of heavy quarks are suppressed.
The state $(D^{*0} \bar D^0)_+$ is therefore much more likely to arise 
from a $c \bar c$ pair that is created at short distances in a 
spin-triplet state than in a spin-singlet state.
As pointed out in Ref.~\cite{Braaten:2004jg}, this implies the 
suppression of the spin-singlet NRQCD matrix elements: 
\begin{subequations}
\begin{eqnarray}
\langle {\cal O}_1^X ({}^1S_0) \rangle  
&\ll&  \langle {\cal O}_1^X ({}^3S_1) \rangle,
\label{O13:triplet}
\\
\langle {\cal O}_8^X ({}^1S_0) \rangle  
&\ll&  \langle {\cal O}_8^X ({}^3S_1) \rangle .
\label{O13:singlet}
\end{eqnarray}
\label{O13}
\end{subequations}

One possible mechanism for the binding of $X(3872)$ that has not 
been excluded is an accidental fine-tuning of the mass of the 
charmonium state $\chi_{c1}(2P)$ to the $D^{*0} \bar D^0$ threshold.
The $c \bar c$ state is transformed by its resonant coupling 
to $D^{*0} \bar D^0$ into a loosely-bound charm-meson molecule
with the properties described in Section~\ref{sec:X3872}.  
In this case, the relative sizes 
of the NRQCD matrix elements for the $X(3872)$ may be governed 
by the velocity-scaling rules for $^3P_1$ charmonium states.
The leading terms in the NRQCD factorization formula are the 
color-singlet $^3P_1$ and color-octet $^3S_1$ terms \cite{Bodwin:1994jh}:
\begin{equation}
\sigma [ X(3872) ] \approx
\hat \sigma [ c \bar c_1 ({}^3P_1) ]  \;
	\langle {\cal O}_1^X ({}^3P_1) \rangle 
+ \hat \sigma [ c \bar c_8 ({}^3S_1) ]  \;
	\langle {\cal O}_8^X ({}^3S_1) \rangle .
\label{fft:chi1}
\end{equation}
The color-singlet parton cross section 
$ \hat \sigma [ c \bar c_1 ({}^3P_1) ]$
and the color-octet NRQCD matrix element 
$\langle {\cal O}_8^X ({}^3S_1) \rangle$
both depend logarithmically on the NRQCD factorization scale 
$\Lambda_{\rm NRQCD}$ in such a way that the dependence cancels 
between the two terms on the right side of Eq.~(\ref{fft:chi1}).
At any specific value of the momentum $\bm{P}$
of the $c \bar c$ pair, the color-singlet parton cross section 
$\hat \sigma [ c \bar c_1 ({}^3P_1) ]$ can be made to vanish 
by adjusting $\Lambda_{\rm NRQCD}$.   
If the production is dominated by regions of $\bm{P}$ in which 
the ratio of $\hat \sigma [ c \bar c_1 ({}^3P_1) ]$
to $\hat \sigma [ c \bar c_8 ({}^3S_1) ]$ is roughly constant,
then $\Lambda_{\rm NRQCD}$ can be chosen so that
the color-octet $^3S_1$ term in the factorization formula 
in Eq.~(\ref{fft:chi1}) dominates.
Thus the S-wave factorization formula in Eq.~(\ref{fft:X2})
may be applicable even if the production of $X(3872)$
is dominated by the formation of $\chi_{c1}(2P)$.

\subsection{Determination of the NRQCD matrix elements}
\label{sec:NRQCD_me}

The NRQCD matrix elements in the NRQCD factorization 
formula in Eq.~(\ref{fft:X2}) are universal.  Once they have been 
determined by fitting experimental results, 
the NRQCD factorization formula can be used to predict 
the inclusive production rates of the $X(3872)$ in other experiments. 
The estimate in Eq.~(\ref{exp_sigma_X-2}) for the prompt cross section 
at the Tevatron provides a single linear
constraint on the products of the matrix elements
and the branching fraction for $X\rightarrow J/\psi \pi^+ \pi^-$.
In order to derive this constraint, we use 
MadOnia~\cite{Artoisenet:2007qm} to evaluate the parton cross sections
appearing in Eq.~(\ref{fft:X2}) at leading order in $\alpha_s$, 
which is order $\alpha_s^3$ for $p_T(c \bar c)>0$.
We identify the 3-momentum $\bm{P}$ of the $X(3872)$ with the 
3-momentum of the $c \bar c$ pair in the center-of-momentum frame 
of the colliding hadrons.
We set the mass of the charm quark to $m_c=1.5$~GeV,
so a $c \bar c$ pair with vanishing relative momentum has mass
3~GeV.  The phase space region for the $c \bar c$ pair is 
$p_T(c \bar c) > 5$~GeV and $|y(c \bar c)| < 0.6$.
We use the parton distribution function Cteq\_6l1~\cite{Pumplin:2002vw}.
The factorization and renormalization scales 
are set equal to the transverse mass $(p_T^2 + 4 m_c^2)^{1/2}$ 
of the charm-quark pair.  
The resulting linear constraint on the NRQCD matrix elements 
from our estimate in Eq.~(\ref{exp_sigma_X-2}) 
for the prompt cross section at the Tevatron is
\begin{eqnarray}
&& {\rm Br}[X\rightarrow J/\psi \pi^+ \pi^-] 
\left( \langle {\cal O}_8^X ({}^3S_1) \rangle
+ 0.159~\langle {\cal O}_8^X ({}^1S_0) \rangle 
+ 0.085~\langle {\cal O}_1^X ({}^1S_0) \rangle \right.
\nonumber
\\
&& 
\hspace{4cm} \left. 
+ 0.00024~\langle {\cal O}_1^X ({}^3S_1) \rangle \right)
= (2.7 \pm 0.6) \times 10^{-4}~{\rm GeV}^3 .
\label{<O>-constraint}
\end{eqnarray}
The error bar comes only from the statistical uncertainty in
the estimate of the prompt cross section 
at the Tevatron in Eq.~(\ref{exp_sigma_X-2}).

Simplifying assumptions can be used to reduce the four independent 
S-wave NRQCD matrix elements in Eq.~(\ref{fft:X2}) to a smaller set.
We will consider three different simplifying assumptions that
reduce the four matrix elements to a single non-perturbative factor,
which we choose to be $\langle \mathcal{O}_8^X (^3S_1) \rangle $.
Our three simplifying assumptions are as follows:

\begin{enumerate}

\item 
\underline{S-wave dominance}.
The $X(3872)$ is equally likely to be formed from any $c \bar c$ pair 
that is created with small relative momentum in an S-wave state,
regardless of the color or spin state of the $c \bar c$ pair. 
The assumptions on the NRQCD matrix elements are
\begin{subequations}
\begin{eqnarray}
\langle {\cal O}_1^X ({}^3S_1) \rangle &=& 
\frac34 \langle {\cal O}_8^X ({}^3S_1) \rangle,
\\
\langle {\cal O}_1^X ({}^1S_0) \rangle &=& 
\frac14 \langle {\cal O}_8^X ({}^3S_1) \rangle,
\\
\langle {\cal O}_8^X ({}^1S_0) \rangle &=& 
\frac13 \langle {\cal O}_8^X ({}^3S_1) \rangle .
\end{eqnarray}
\label{Sdominance}
\end{subequations}
This is the same pattern of matrix elements as in the color-evaporation model 
for quarkonium production \cite{Bodwin:2005hm}. 
The linear constraint in Eq.~(\ref{<O>-constraint}) reduces to
${\rm Br}~\langle {\cal O}_8^X ({}^3S_1) \rangle
\approx 2.5 \times 10^{-4}~{\rm GeV}^3$.

\item 
\underline{Spin-triplet dominance}.
The $X(3872)$ can be formed only from a $c \bar c$ pair that is created 
with small relative momentum in a spin-triplet S-wave state,
but it is equally likely to be formed if the $c \bar c$ pair
is in a color-singlet or color-octet state.
The assumptions on the NRQCD matrix elements are
\begin{subequations}
\begin{eqnarray}
\langle {\cal O}_1^X ({}^3S_1) \rangle &=& 
\frac34 \langle {\cal O}_8^X ({}^3S_1) \rangle,
\\
\langle {\cal O}_1^X ({}^1S_0) \rangle &=& 
\langle {\cal O}_8^X ({}^1S_0) \rangle = 0.
\end{eqnarray}
\label{S3dominance}
\end{subequations}
The linear constraint in Eq.~(\ref{<O>-constraint}) reduces to
${\rm Br}~\langle {\cal O}_8^X ({}^3S_1) \rangle
\approx 2.7 \times 10^{-4}~{\rm GeV}^3$.

\item 
\underline{Color-octet $^3S_1$ dominance}.
The $X(3872)$ can be formed only from a $c \bar c$ pair that is created 
with small relative momentum in a color-octet ${}^3S_1$ state.
The assumptions on the NRQCD matrix elements are
\begin{eqnarray}
  \langle {\cal O}_1^X ({}^3S_1) \rangle 
= \langle {\cal O}_1^X ({}^1S_0) \rangle 
= \langle {\cal O}_8^X ({}^1S_0) \rangle  =  0 .
\label{S83dominance}
\end{eqnarray}
This simplifying assumption was proposed by Braaten,
who suggested that it might give a good approximation 
to the inclusive production rate of the $X(3872)$ 
in high-energy hadron collisions~\cite{Braaten:2004jg}.
The linear constraint in Eq.~(\ref{<O>-constraint}) reduces to
${\rm Br}~\langle {\cal O}_8^X ({}^3S_1) \rangle
\approx 2.7 \times 10^{-4}~{\rm GeV}^3$.

\end{enumerate}

\begin{figure}[t]
\includegraphics*[width=8cm,angle=0,clip=true]{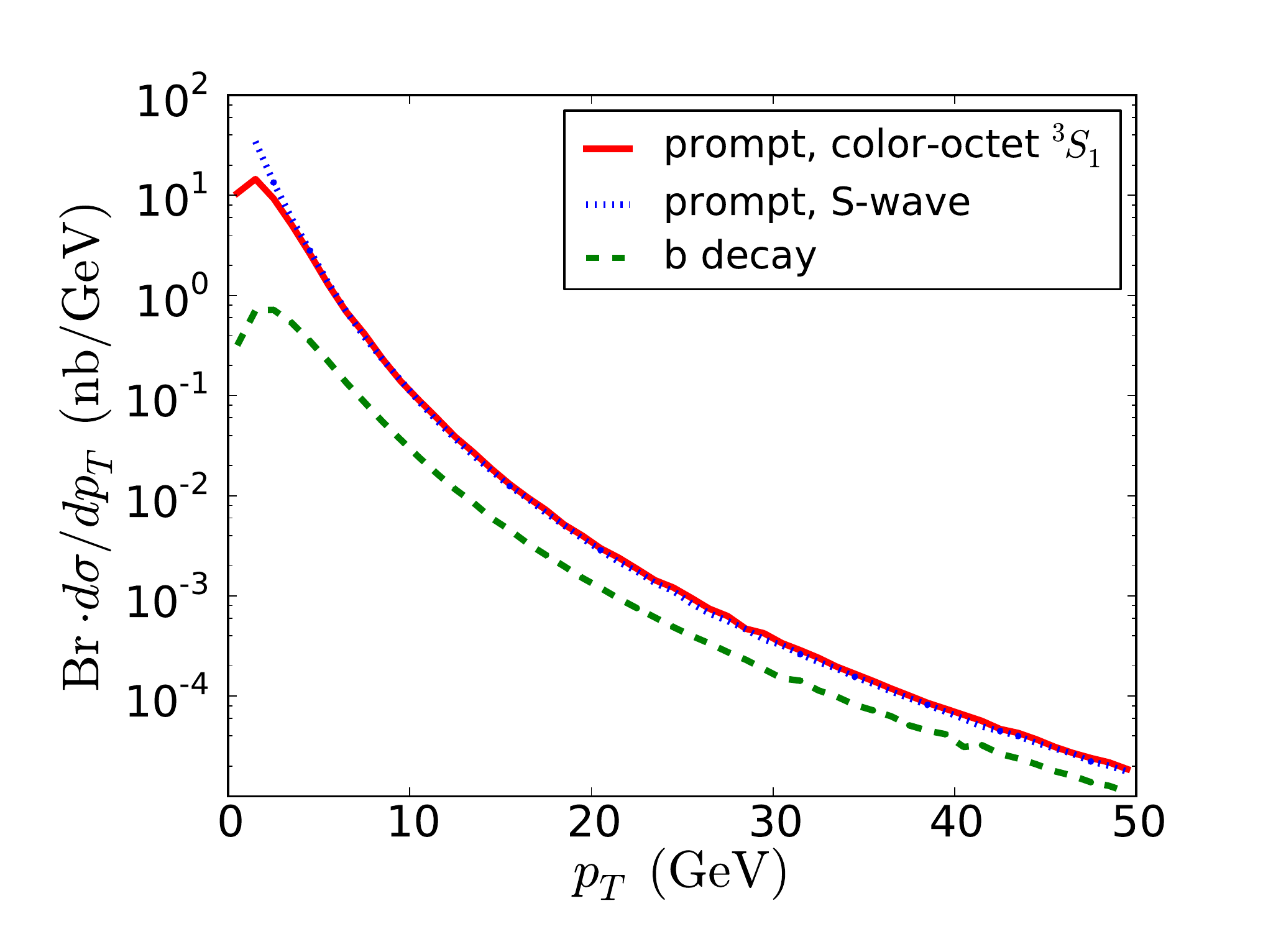}
\includegraphics*[width=8cm,angle=0,clip=true]{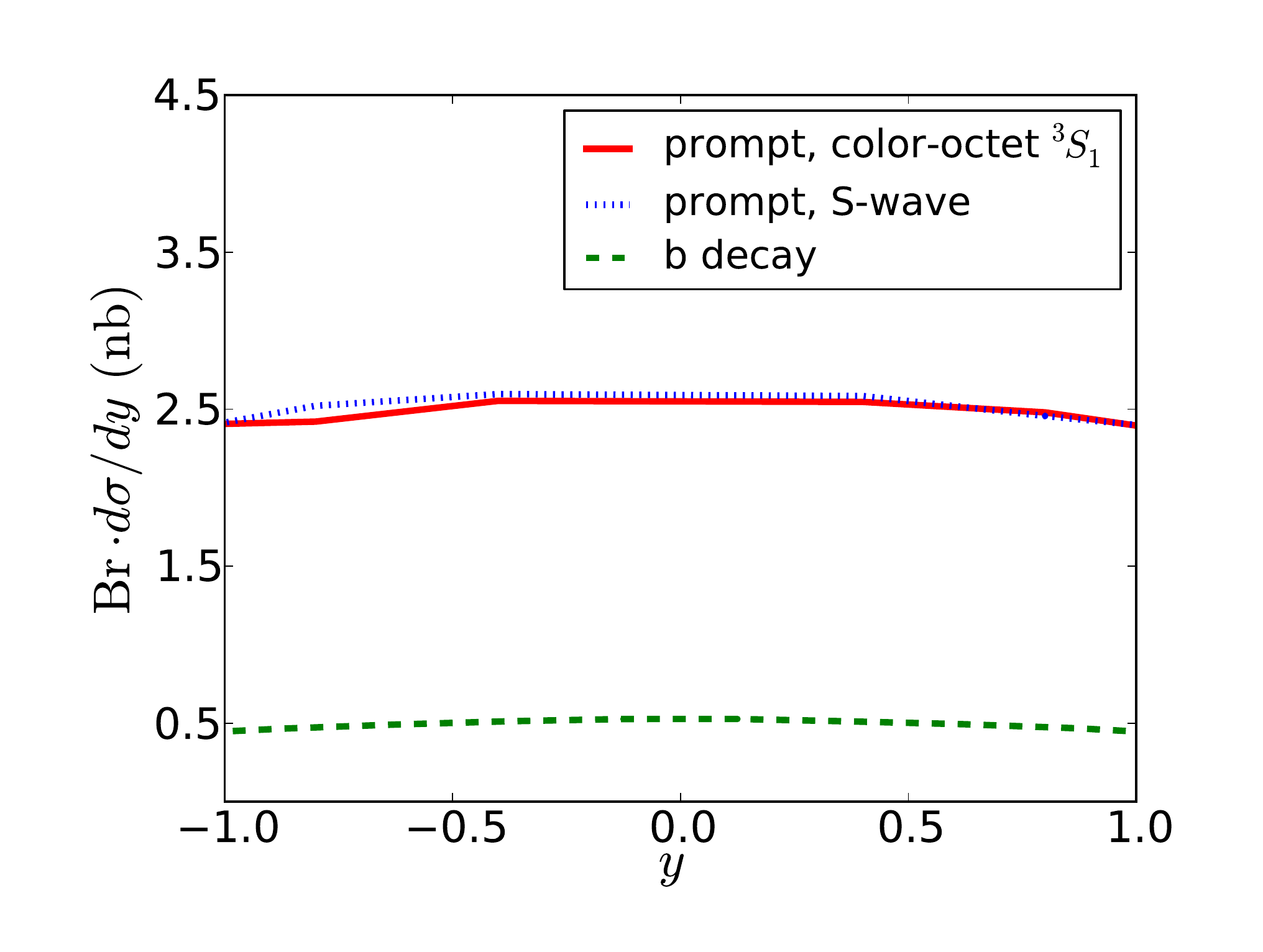}
\vspace*{0.0cm}
\caption{Cross sections for 
$X(3872) \rightarrow J/\psi \pi^+ \pi^-$ 
in $p \bar p$ collisions at $\sqrt{s} = 1.96$~TeV.
The graphs are the transverse momentum ($p_T$) distribution 
for $|y| < 0.6$ (left panel) and  the rapidity
($y$) distribution for $p_T> 5$~GeV (right panel).
The curves are for prompt production assuming color-octet $^3S_1$ 
dominance (solid) or S-wave dominance (dotted)
and for production from $b$-hadron decay (dashed).
}
\label{figTev}
\end{figure}

Having obtained estimates of the NRQCD matrix elements 
for each of our three simplifying assumptions, we can now predict 
the differential cross sections for prompt production of 
$X \to J/\psi \pi^+ \pi^-$ at the Tevatron.  The distributions 
of the transverse momentum and the rapidity
are shown in Fig.~\ref{figTev}.
The solid and dotted curves correspond to 
color-octet $^3S_1$ dominance and S-wave dominance, respectively.
The curves for spin-triplet dominance cannot be distinguished 
from those for color-octet $^3S_1$ dominance.
For $p_T$ larger than about 4~GeV, our three simplifying
assumptions on the NRQCD matrix elements
lead to the same differential cross section.
The reason for this is that the order-$\alpha_s^3$ parton cross section 
$\hat \sigma [ c \bar c_8 ({}^3S_1) ]$ dominates over that for 
the other three S-wave channels at large $p_T$.
In the color-octet $^3S_1$ channel, the order-$\alpha_s^3$ 
parton cross section includes a fragmentation contribution 
that decreases asymptotically as 
$d\hat \sigma/dp_T^2 \sim 1/p_T^4$.  For the other three channels,
the order-$\alpha_s^3$ parton cross section decreases more rapidly
by a factor of $1/p_T^2$ or $1/p_T^4$.  They receive fragmentation 
contributions only at order $\alpha_s^4$ or $\alpha_s^5$.
Thus the differences between our three simplifying 
assumptions might be greater at large $p_T$ 
if higher order corrections in $\alpha_s$ were included.  
These corrections are beyond the scope of this paper.
As $p_T$ decreases below 4~GeV, the differential cross section
for S-wave dominance continues to increase while that 
from the color-octet $^3S_1$ dominance reaches a maximum 
and then begins to decrease.
The reason the S-wave dominance cross section increases
is that the order-$\alpha_s^3$ parton cross sections 
in the color-singlet and color-octet ${}^1S_0$ channels
diverge as $p_T \rightarrow 0$.  
At very low transverse momentum, a fixed-order 
calculation for this channel is not reliable, 
and one has to resum higher order corrections in $\alpha_s$
associated with soft-gluon radiation from the colliding partons.  
This resummation is beyond the scope of this paper.

In the differential cross sections in Fig.~\ref{figTev},
there is a 23\% uncertainty in the normalization 
from the NRQCD matrix element.  There are additional theoretical 
uncertainties associated with the NRQCD factorization formalism.
The normalizations of the parton cross sections are sensitive to the value 
of $m_c$ and the renormalization and factorization 
scales.  However the shapes of the differential cross section 
are much less sensitive.  Thus variations in these parameters 
can be largely compensated by changes in the value of  
$\langle {\cal O}_8^X ({}^3S_1) \rangle$.

\subsection{Production from $\bm{b}$-hadron decay}
\label{sec:bdecay}

Beside prompt production, the other source of $X(3872)$
in a hadron collider is the feed-down from decays of $b$ hadrons.
At the Tevatron, the rate from $b$-hadron decays is smaller than 
the prompt rate but nevertheless significant. 
In the sample of $X(3872) \rightarrow J/\psi \pi^+ \pi^-$ events 
studied by the CDF Collaboration in Ref.~\cite{Bauer:2004bc}, 
the fraction of events from decays of $b$ hadrons was 
$16.1 \pm 4.9 \pm 2.0 \%$.

To predict the differential cross section for $X(3872)$
from $b$-hadron decays at the Tevatron, we follow a procedure 
similar to the one used in 
Ref.~\cite{Cacciari:2003uh} to calculate the momentum distribution 
of $J/\psi$ from $b$-hadron decays in high energy collisions.
The differential cross section for $X(3872)$ is the convolution 
of the cross section for producing a $b$ quark, a fragmention function for
its hadronization into a $b$ hadron, and the momentum distribution 
of the $X(3872)$ from the decay of the $b$ hadron.
The Monte Carlo program MCFM~\cite{Campbell:1999ah} is used
to generate the momentum 
spectrum of $b$ quarks at next-to-leading order in $\alpha_s$. 
We set the mass of the $b$ quark to $m_b=4.75$ GeV.
We use the Cteq6\_m parton distributions~\cite{Pumplin:2002vw}, 
with the factorization and renormalization scales set equal to the 
transverse mass $(p_T^2 + m_b^2)^{1/2}$ of the $b$ quark. 
We use the Kartvelishvili fragmentation function~\cite{Kartvelishvili:1977pi} 
with exponent $\alpha = 29.1$ to describe the 
hadronization of the $b$ quark into a $b$ hadron.
There are several ways in which we depart from the procedure used in
Ref.~\cite{Cacciari:2003uh}.  We
do not resum the logarithms of $m_b/p_T$ at large transverse momentum.
To approximate the momentum distribution of the $X(3872)$ from the 
decay of the $b$ hadron, we take the $b$ hadron to be a $B$ meson 
and we decay it into an $X$ plus a kaon according to an isotropic 
distribution in the rest frame of the $B$ meson.
To normalize the differential cross section for $X(3872)$ 
from $b$-hadron decays, we use the estimated cross section
for $X(3872) \rightarrow J/\psi \pi^+ \pi^-$ at the Tevatron 
given in Eq.~(\ref{exp_sigma_b_X-2}).  The normalizing factor 
can be interpreted as the product of the inclusive branching fraction
for a $b$ quark to decay into $X(3872)$ and the branching fraction
for $X \rightarrow J/\psi \pi^+ \pi^-$:
\begin{equation}
{\rm Br}[b \rightarrow X(3872) + \textrm{any}] \, 
{\rm Br}[X \rightarrow J/\psi \pi^+ \pi^-]
=(1.9 \pm 0.8)\times 10^{-4} .
\label{BrBr}
\end{equation}
The error bar comes only from the statistical error in the estimate of 
the $b$-decay cross section at the Tevatron in Eq.~(\ref{exp_sigma_b_X-2}).

Having determined the normalizing factor in Eq.~(\ref{BrBr}),
we can predict the differential cross sections for producing
$X\rightarrow J/\psi \pi^+ \pi^-$ from $b$ decays at the 
Tevatron.  The predictions are shown as dashed curves in Fig.~\ref{figTev}.
The shapes of the distributions in $p_T$ and $y$ are predicted to be 
similar to those for prompt production by the color-octet $^3S_1$ 
mechanism except at low transverse momentum.  
The fraction of $X(3872)$ from $b$-hadron decay 
is predicted to increase from about 12\% at $p_T = 5$~GeV to about
37\% at $p_T = 50$~GeV.

\section{Production at the LHC}
\label{sec:LHC}

In this section, we predict the differential cross sections for 
the production of the $X(3872)$ at the Large Hadron Collider (LHC)
from both prompt QCD mechanisms and the decays of $b$ hadrons.
We consider the production of $X(3872)$ in the phase space regions of 
the ATLAS and CMS detectors and the LHCb detector.

\begin{figure}[t]
\includegraphics*[width=8cm,angle=0,clip=true]{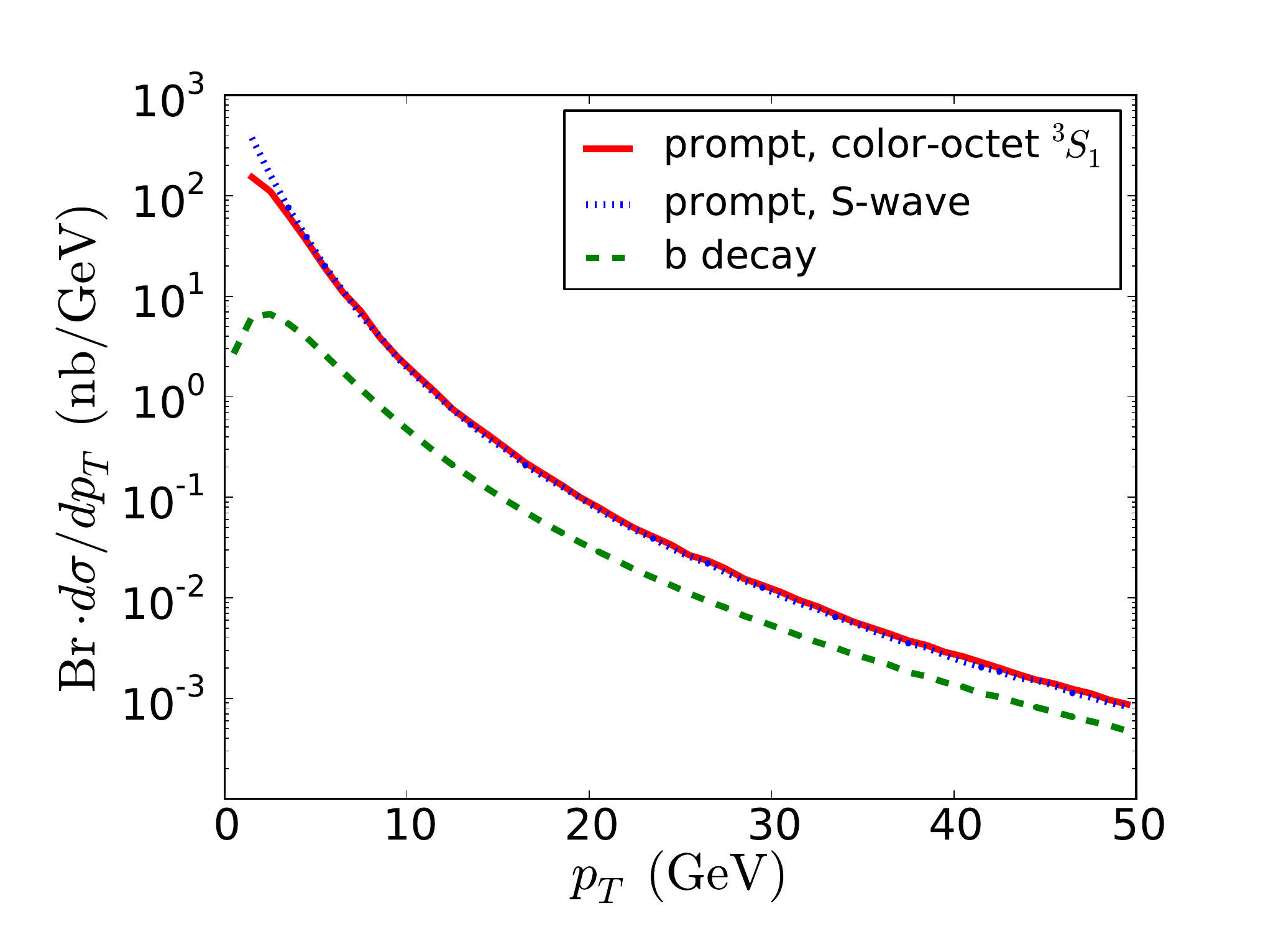}
\includegraphics*[width=8cm,angle=0,clip=true]{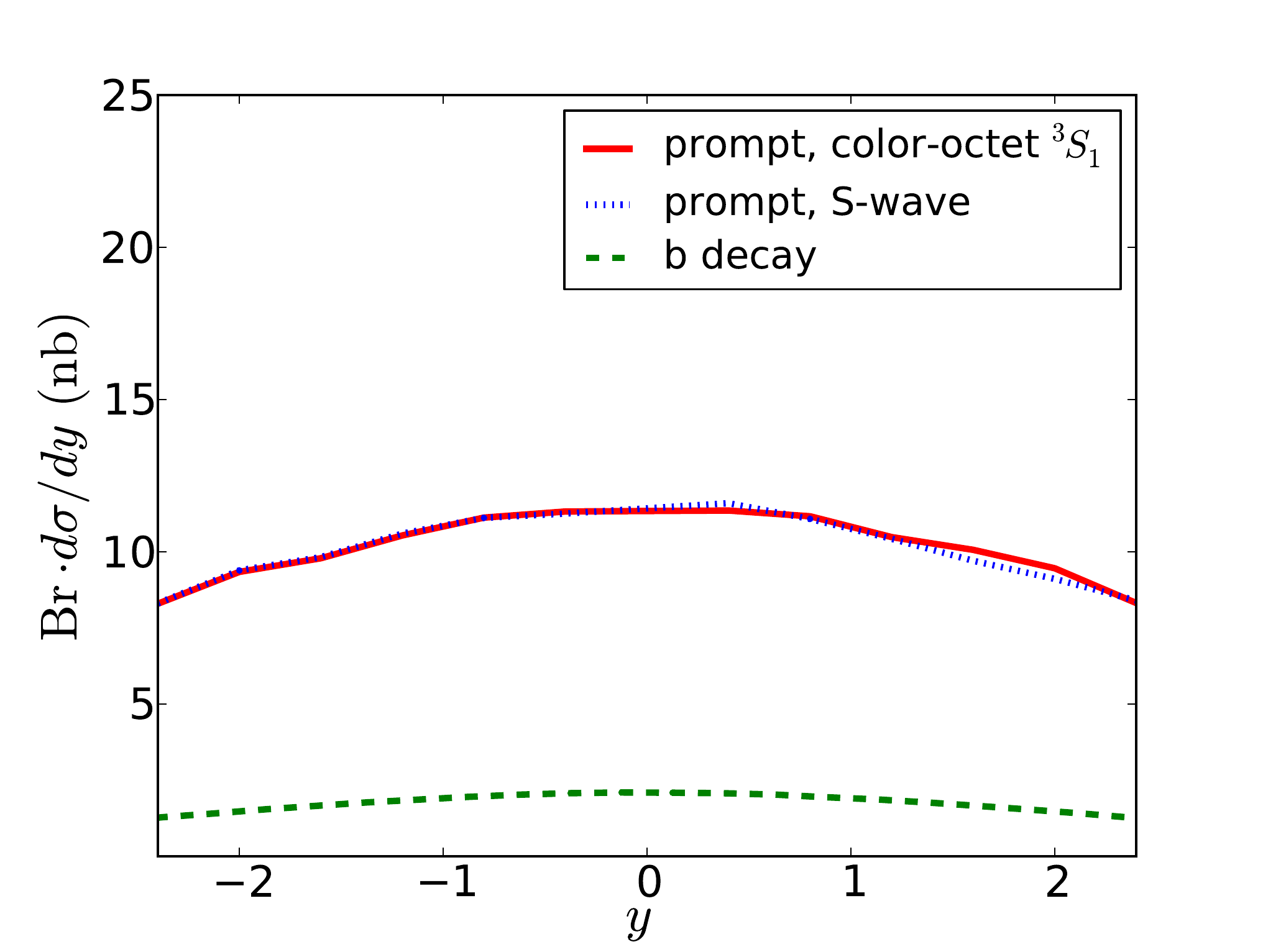}
\vspace*{0.0cm}
\caption{Cross sections for 
$X (3872)\rightarrow J/\psi \pi^+ \pi^-$ 
in $p p$ collisions at $\sqrt{s} = 7$~TeV.
The graphs are the transverse momentum ($p_T$) distribution 
for $|y| < 2.4$ (left panel) and  the rapidity
($y$) distribution for $p_T> 5$~GeV (right panel).
The curves are for prompt production assuming color-octet $^3S_1$ 
dominance (solid) or S-wave dominance (dotted)
and for production from $b$-hadron decay (dashed).
}
\label{fig5}
\end{figure}

To predict the differential cross sections for production 
of $X(3872)$ at the LHC, we use the same methods that were 
used to calculate the differential cross sections at the Tevatron 
shown in Fig.~\ref{figTev}.  We calculate the prompt cross section
for $X(3872)$ using the NRQCD factorization formula in Eq.~(\ref{fft:X2}).
We consider the three  simplifying assumptions 
for the NRQCD matrix elements introduced in 
Section~\ref{sec:NRQCD_me}:
S-wave dominance defined by Eqs.~(\ref{Sdominance}),
spin-triplet dominance defined by Eqs.~(\ref{S3dominance}),
and color-octet $^3S_1$ dominance defined by Eq.~(\ref{S83dominance}).
The normalization of the prompt cross section is determined by the 
linear constraint in Eq.~(\ref{<O>-constraint}).
We calculate the $c \bar c$ cross sections using the same parameters 
as in Section~\ref{sec:NRQCD_me}.
We calculate the cross section for $X(3872)$ from $b$-hadron decays 
using the method described in Section~\ref{sec:bdecay}.
The normalization of the $b$ decay cross section is determined 
by the product of branching fractions given in Eq.~(\ref{BrBr}).

We first consider the production of $X(3872)$ 
in the phase-space region covered by the CMS and ATLAS detectors. 
In Fig.~\ref{fig5}, we show the $p_T$ distributions 
integrated over the rapidity range $|y|<2.4$ 
and the rapidity distributions integrated over the region $p_T>5$~GeV. 
The qualitative behavior of these curves is 
similar to those for the Tevatron in Fig.~\ref{figTev}.
The differences in the prompt cross section from
different simplifying assumptions for the NRQCD matrix elements 
are negligible for $p_T$ greater than about 5~GeV.
The prompt and $b$-decay cross sections integrated over the 
region $p_T>5$ GeV and $|y|<2.4$ are predicted to be about 49~nb
and 8.2~nb, respectively. 
The fraction of $X(3872)$ events from $b$-hadron decay is predicted to
increase from 10\% at $p_T=5$~GeV to 35\% at $p_T=50$~GeV. 
However there are large uncertainties in the predictions.
The statistical uncertainties in the normalizing factors
are 23\% for the prompt cross section 
and 39\% for the $b$-decay cross section.
There are also additional theoretical errors from the masses of the 
charm and bottom quarks and from the choices of the factorization 
and renormalization scales. 
Although these uncertainties are large, they will partly cancel 
between the extraction of the normalizing factors from the Tevatron 
data and the calculation of the cross sections at the LHC.

\begin{figure}[t]
\includegraphics*[width=8cm,angle=0,clip=true]{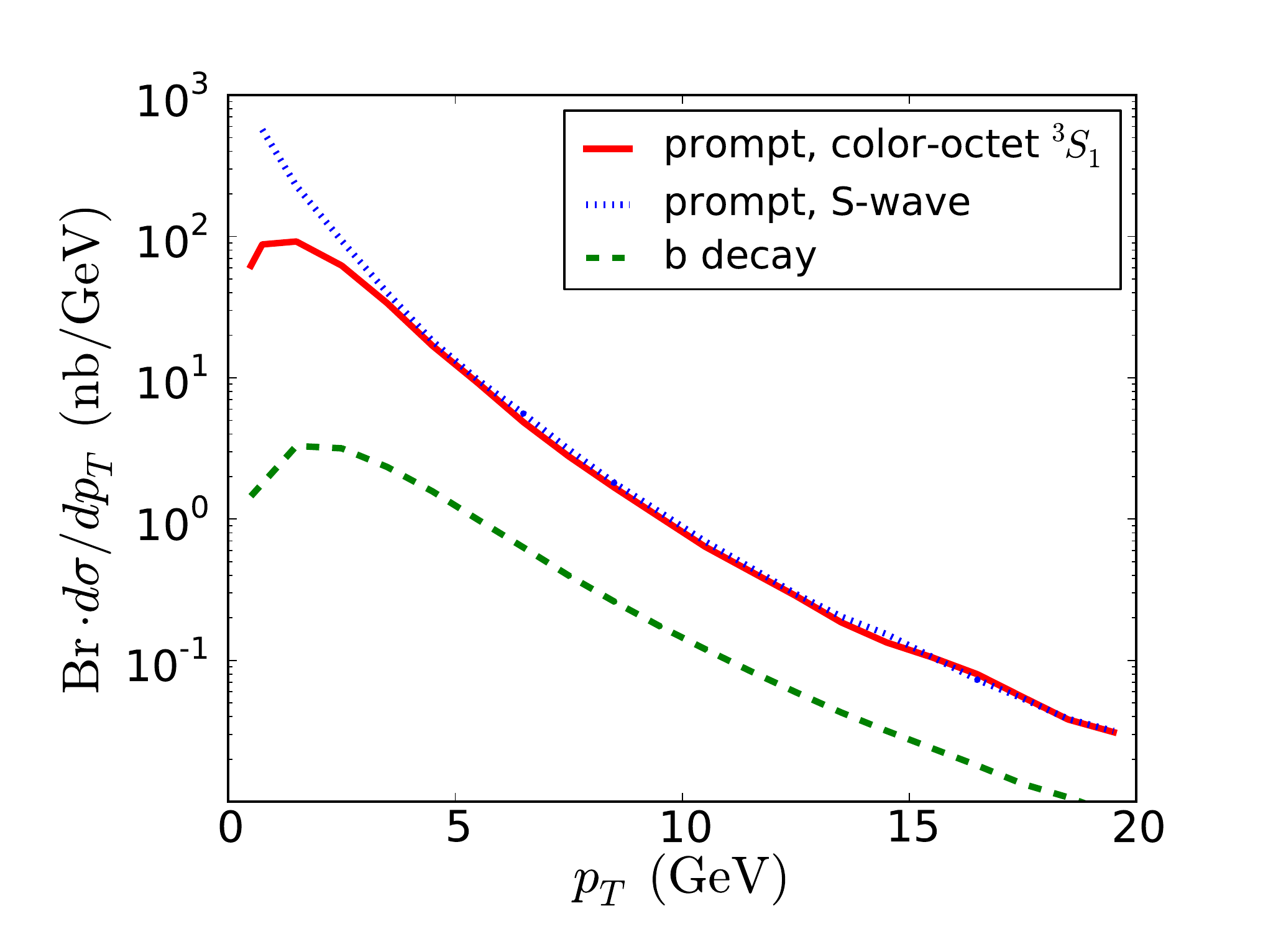}
\includegraphics*[width=8cm,angle=0,clip=true]{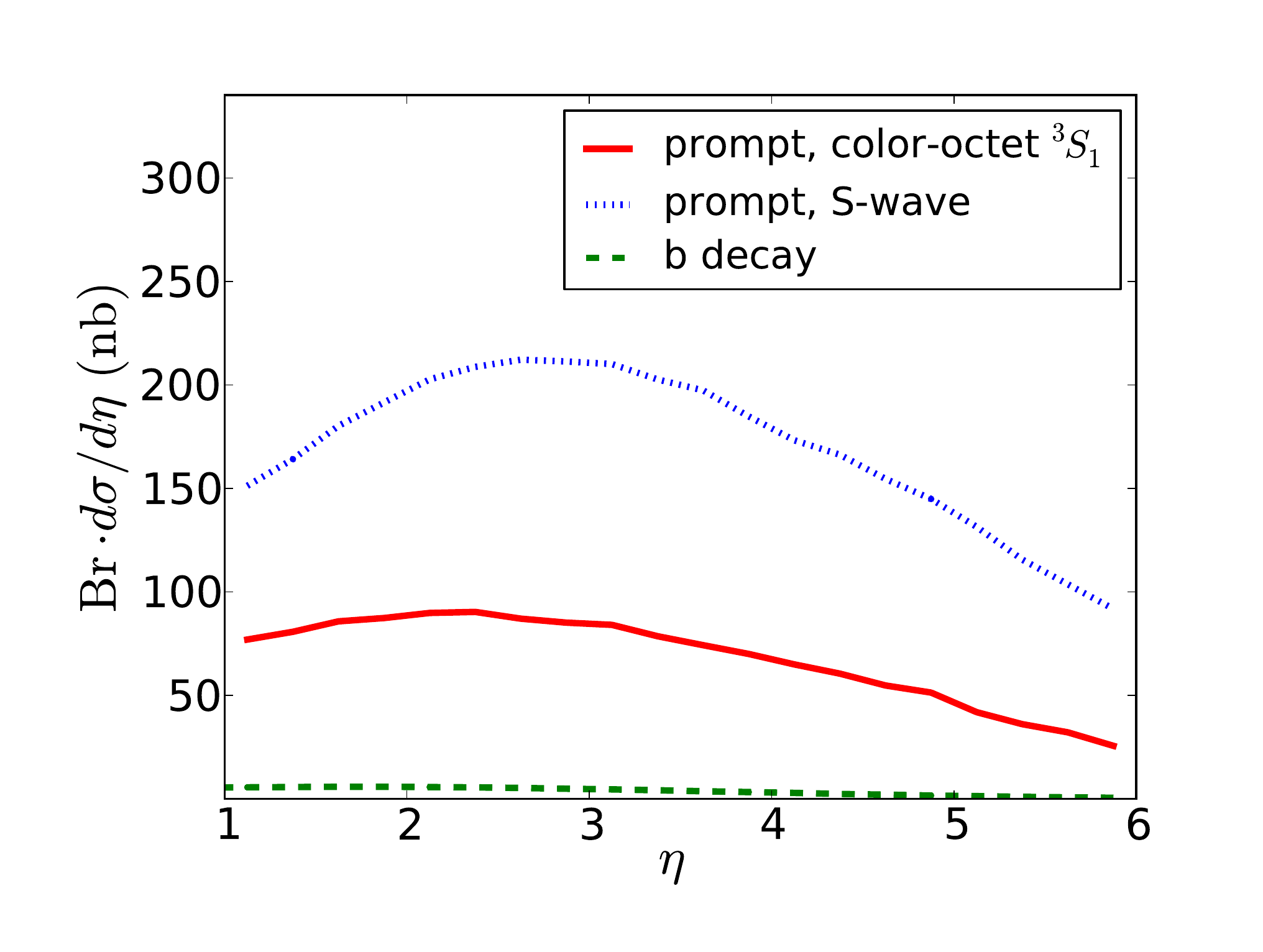}
\vspace*{0.0cm}
\caption{
Cross sections for $X\rightarrow J/\psi \pi^+ \pi^-$ 
in $p p$ collisions at $\sqrt{s} = 7$~TeV.
The graphs are the transverse momentum ($p_T$) distribution 
for $1.6 < \eta < 5.3$ (left panel) and 
the pseudorapidity ($\eta$) distribution for $p_T> 0.5$~GeV (right panel).
The curves are for prompt production assuming color-octet $^3S_1$ 
dominance (solid) or S-wave dominance (dotted)
and for production from $b$-hadron decay (dashed).}
\label{fig6}
\end{figure}

We now move on to the production of $X(3872)$ in the LHCb experiment, 
which is a forward detector in the pseudorapidity region $1.6<\eta<5.3$.  
In Fig~\ref{fig6}, we show the $p_T$ distributions integrated 
over the pseudorapidity range $1.6 < \eta < 5.3$ and the 
pseudorapidity distributions integrated over the region $p_T>0.5$~GeV.
The differences in the prompt cross section from
different simplifying assumptions for the NRQCD matrix elements 
are small for $p_T$ greater than about 5~GeV, but they increase at
smaller values of $p_T$. The cross section for S-wave dominance
is larger than that for color-octet $^3S_1$ dominance by a factor of
2.5 at $p_T=1.5$~GeV and 6.5 at $p_T=0.75$~GeV.  This leads to the 
large difference in the normalization of the rapidity distributions 
for these two cases in Fig.~\ref{fig6}.
The prompt cross section integrated over the 
region $p_T>0.5$ GeV and $1.6<\eta<5.3$ is predicted to be 
about 270~nb and 688~nb in the cases of color-octet $^3S_1$ dominance
and S-wave dominance, respectively. 
The $b$-decay cross section integrated over the same region
is predicted to be 14~nb. 
Assuming color-octet $^3S_1$ dominance, the fraction 
of $X(3872)$ events from $b$-hadron decay is predicted to
increase from 2.2\% at $p_T=0.5$~GeV to 9.0\% at $p_T=5$~GeV and to
20\% at $p_T=20$~GeV. 
There are large uncertainties in these predictions,
especially those that involve production at small $p_T$.

\section{Summary}
\label{sec:summary}

We have analyzed the production of the $X(3872)$ at the Tevatron. 
We showed that the prompt production rate observed at the Tevatron 
is compatible with the identification of $X(3872)$ as a loosely-bound
charm meson molecule. 
In Ref.~\cite{Bignamini:2009sk}, the authors found that the production 
rate at the Tevatron was too large for a loosely-bound charm-meson 
molecule by orders of magnitude.  The error in their analysis 
was that they did not take into account effects of rescattering 
of the charm mesons that are specific to an S-wave threshold resonance.

We have predicted the differential cross sections for the $X(3872)$ 
at the LHC from both prompt QCD mechanisms and the decays of $b$ hadrons.
The prompt cross section for $X(3872)$ was calculated 
by using the NRQCD factorization formula.  
We used simplifying assumptions to reduce 
the nonperturbative factors to a single NRQCD matrix element
$\langle {\cal O}_8^X ({}^3S_1) \rangle$, which was determined 
from an estimate of the prompt cross section at the Tevatron.
The cross section for $X(3872)$ from $b$-hadron decay was calculated 
using a method that gives the correct differential cross section 
for $J/\psi$ from $b$-hadron decay.
The normalizing factor was determined 
from an estimate of the $b$-decay cross section at the Tevatron.

For the ATLAS and CMS detectors, we predict the production rate 
to be more than an order of magnitude larger than for the CDF detector at the
Tevatron, given the same cut on the $p_T$ of the $X$
and the same integrated luminosity.  For the LHCb detector,
where a much lower $p_T$ cut is possible, the production rate 
could be larger by another order of magnitude. 
Thus the LHC experiments should be able to collect 
very large data samples of $X(3872)$.

The large data samples of $X(3872)$ at the LHC
will allow precise measurements of various properties of this remarkable
hadron. It should be possible to measure the branching ratios for decays
of $X(3872)$ with a good signature, including $J/\psi \pi^+ \pi^-$, 
 $J/\psi \gamma$, and $\psi(2S) \gamma$.
It may also be possible to observe for the first time some rare decay
modes of the $X(3872)$, such as $\chi_{cJ}(1P)\pi^+ \pi^-$~\cite{Dubynskiy:2007tj,Fleming:2008yn}.
It should be possible to improve on the measurement of the 
binding energy of $X(3872)$ by the CDF Collaboration \cite{Aaltonen:2009vj}. 
It may even be possible to distinguish the universal line shape 
of the $X(3872)$ \cite{Braaten:2007dw,Braaten:2007ft}
from that of a conventional Breit-Wigner resonance.
In conclusion, the experiments at the LHC are sure to add significantly
to our understanding of the $X(3872)$.

\begin{acknowledgments}
This research was supported in part by the Department of Energy 
under grant DE-FG02-05ER15715.
\end{acknowledgments}

\begin{appendix}

\end{appendix}


\begin{thebibliography}{99}

\bibitem{Choi:2003ue}
  S.K.~Choi {\it et al.}  [Belle Collaboration],
  Phys.\ Rev.\ Lett.\  {\bf 91}, 262001 (2003)
  [arXiv:hep-ex/0309032].

\bibitem{Acosta:2003zx}
  D.~E.~Acosta {\it et al.}  [CDF II Collaboration],
  Phys.\ Rev.\ Lett.\  {\bf 93}, 072001 (2004)
  [arXiv:hep-ex/0312021].
  
\bibitem{Bauer:2004bc}
  G.~Bauer  [CDF II Collaboration],
  Int.\ J.\ Mod.\ Phys.\  A {\bf 20}, 3765 (2005)
  [arXiv:hep-ex/0409052].

\bibitem{Voloshin:2003nt}
  M.~B.~Voloshin,
  Phys.\ Lett.\  B {\bf 579}, 316 (2004)
  [arXiv:hep-ph/0309307].

\bibitem{Braaten:2003he}
  E.~Braaten and M.~Kusunoki,
  Phys.\ Rev.\  D {\bf 69}, 074005 (2004)
  [arXiv:hep-ph/0311147].

\bibitem{Bignamini:2009sk}
  C.~Bignamini, B.~Grinstein, F.~Piccinini, A.~D.~Polosa and C.~Sabelli,
  arXiv:0906.0882 [hep-ph].

\bibitem{Abulencia:2006ma}
  A.~Abulencia {\it et al.}  [CDF Collaboration],
  Phys.\ Rev.\ Lett.\  {\bf 98}, 132002 (2007)
  [arXiv:hep-ex/0612053].
  
\bibitem{Aaltonen:2009vj}
  T.~Aaltonen {\it et al.}  [CDF Collaboration],
  arXiv:0906.5218 [hep-ex].
 
\bibitem{Braaten:2004jg}
  E.~Braaten,
  Phys.\ Rev.\  D {\bf 73}, 011501 (2006)
  [arXiv:hep-ph/0408230].

\bibitem{Abe:2005ix}
  K.~Abe {\it et al.},
  arXiv:hep-ex/0505037.

\bibitem{Aubert:2006aj}
  B.~Aubert {\it et al.}  [BABAR Collaboration],
  Phys.\ Rev.\  D {\bf 74}, 071101 (2006)
  [arXiv:hep-ex/0607050].

\bibitem{Abe:2005iya}
  K.~Abe {\it et al.},
  arXiv:hep-ex/0505038.

\bibitem{Gokhroo:2006bt}
  G.~Gokhroo {\it et al.},
  Phys.\ Rev.\ Lett.\  {\bf 97}, 162002 (2006).
  [arXiv:hep-ex/0606055].

\bibitem{Babar:2008rn}
  B.~Fulsom {\it et al.}  [BABAR Collaboration],
  arXiv:0809.0042 [hep-ex].

\bibitem{Aubert:2008gu}
  B.~Aubert {\it et al.}  [BABAR Collaboration],
  Phys.\ Rev.\  D {\bf 77}, 111101 (2008)
  [arXiv:0803.2838 [hep-ex]].

\bibitem{Belle:2008te}
 I.~Adachi {\it et al.}  [Belle Collaboration],
  arXiv:0809.1224 [hep-ex].
 
\bibitem{Braaten:2004rn}
  E.~Braaten and H.~W.~Hammer,
  Phys.\ Rept.\  {\bf 428}, 259 (2006)
  [arXiv:cond-mat/0410417].

\bibitem{Braaten:2005jj}
  E.~Braaten and M.~Kusunoki,
  Phys.\ Rev.\  D {\bf 72}, 014012 (2005)
  [arXiv:hep-ph/0506087].

\bibitem{Migdal-Watson}
K.M.~Watson, Phys.\ Rev.\  D {\bf 88}, 1163 (1952);
A.B.~Migdal, JETP {\bf 1}, 2 (1955).

\bibitem{Abazov:2004kp}
  V.~M.~Abazov {\it et al.}  [D0 Collaboration],
  Phys.\ Rev.\ Lett.\  {\bf 93}, 162002 (2004)
  [arXiv:hep-ex/0405004].

\bibitem{Aaltonen:2009dm}
  T.~Aaltonen {\it et al.}  [CDF Collaboration],
  Phys.\ Rev.\  D {\bf 80}, 031103 (2009)
  [arXiv:0905.1982 [hep-ex]].

\bibitem{Amsler:2008zzb}
  C.~Amsler {\it et al.}  [Particle Data Group],
  Phys.\ Lett.\  B {\bf 667}, 1 (2008).

\bibitem{Aubert:2005vi}
  B.~Aubert {\it et al.}  [BABAR Collaboration],
  Phys.\ Rev.\ Lett.\  {\bf 96}, 052002 (2006)
  [arXiv:hep-ex/0510070].

\bibitem{:2008rn}
  B.~Aubert {\it et al.}  [BABAR Collaboration],
  Phys.\ Rev.\ Lett.\  {\bf 102}, 132001 (2009)
  [arXiv:0809.0042 [hep-ex]].

\bibitem{Sjostrand:2006za}
  T.~Sjostrand, S.~Mrenna and P.~Skands,
  JHEP {\bf 0605}, 026 (2006)
  [arXiv:hep-ph/0603175].

\bibitem{Corcella:2000bw}
  G.~Corcella {\it et al.},
  JHEP {\bf 0101}, 010 (2001)
  [arXiv:hep-ph/0011363].

\bibitem{Alwall:2007st}
  J.~Alwall {\it et al.},
  JHEP {\bf 0709}, 028 (2007)
  [arXiv:0706.2334 [hep-ph]].

\bibitem{Reisert:2007zza}
  B.~Reisert  [CDF Collaboration],
  Nucl.\ Phys.\ Proc.\ Suppl.\  {\bf 170}, 243 (2007).

\bibitem{Bodwin:1994jh}
  G.~T.~Bodwin, E.~Braaten and G.~P.~Lepage,
  Phys.\ Rev.\  D {\bf 51}, 1125 (1995)
  [Erratum-ibid.\  D {\bf 55}, 5853 (1997)]
  [arXiv:hep-ph/9407339].

\bibitem{Voloshin:2004mh}
  M.~B.~Voloshin,
  Phys.\ Lett.\  B {\bf 604}, 69 (2004)
  [arXiv:hep-ph/0408321].

\bibitem{Artoisenet:2007qm}
  P.~Artoisenet, F.~Maltoni and T.~Stelzer,
  JHEP {\bf 0802}, 102 (2008)
  [arXiv:0712.2770 [hep-ph]].

\bibitem{Pumplin:2002vw}
  J.~Pumplin, D.~R.~Stump, J.~Huston, H.~L.~Lai, P.~M.~Nadolsky and W.~K.~Tung,
  JHEP {\bf 0207}, 012 (2002)
  [arXiv:hep-ph/0201195].

\bibitem{Bodwin:2005hm}
  G.~T.~Bodwin, E.~Braaten and J.~Lee,
  Phys.\ Rev.\  D {\bf 72}, 014004 (2005)
  [arXiv:hep-ph/0504014].

\bibitem{Cacciari:2003uh}
  M.~Cacciari, S.~Frixione, M.~L.~Mangano, P.~Nason and G.~Ridolfi,
  JHEP {\bf 0407}, 033 (2004)
  [arXiv:hep-ph/0312132].
 
\bibitem{Campbell:1999ah}
  J.~M.~Campbell and R.~K.~Ellis,
  Phys.\ Rev.\  D {\bf 60}, 113006 (1999)
  [arXiv:hep-ph/9905386].

\bibitem{Kartvelishvili:1977pi}
  V.~G.~Kartvelishvili, A.~K.~Likhoded and V.~A.~Petrov,
  Phys.\ Lett.\  B {\bf 78}, 615 (1978).

\bibitem{Dubynskiy:2007tj}
  S.~Dubynskiy and M.~B.~Voloshin,
  Phys.\ Rev.\  D {\bf 77}, 014013 (2008)
  [arXiv:0709.4474 [hep-ph]].

\bibitem{Fleming:2008yn}
  S.~Fleming and T.~Mehen,
  Phys.\ Rev.\  D {\bf 78}, 094019 (2008)
  [arXiv:0807.2674 [hep-ph]].


\bibitem{Braaten:2007dw}
  E.~Braaten and M.~Lu,
  Phys.\ Rev.\  D {\bf 76}, 094028 (2007)
  [arXiv:0709.2697 [hep-ph]].

\bibitem{Braaten:2007ft}
  E.~Braaten and M.~Lu,
  Phys.\ Rev.\  D {\bf 77}, 014029 (2008)
  [arXiv:0710.5482 [hep-ph]].

\end{thebibliography}
\end{document}